\newcommand{\NWtarget}[2]{#2}
\newcommand{\NWlink}[2]{#2}

\newcommand{\NWtxtMacroRefIn}{Macro referenced in}

\newcommand{\NWtxtFileDefBy}{File defined by}
\newcommand{\NWsep}{${\diamond}$}
\documentclass[titlepage,12pt]{article}   

\newif\ifshowresults
\showresultstrue

\newif\ifreleased
\releasedtrue

\usepackage{url}
\usepackage[matrix,arrow,curve]{xy}
\usepackage{graphicx}
\usepackage{alltt}
\usepackage{amsmath}
\usepackage{amsthm}
\theoremstyle{definition}

\newtheorem{role}{Role}
\newtheorem{scenario}{Scenario}

\newcommand{\nterm}[1]{\ensuremath{\langle\mathit{#1}\rangle}}
\newcommand{\nterms}[1]{\ensuremath{\nterm{#1}^\ast}}
\newcommand{\ntermp}[1]{\ensuremath{\nterm{#1}^+}}
\newcommand{\ntermo}[1]{#1$^?$}
\newcommand{\enc}[2]{\{\!|#1|\!\}_{#2}}
\newcommand{\hash}[1]{\#(#1)}
\newcommand{\inbnd}{\mathord -}
\newcommand{\outbnd}{\mathord +}
\newcommand{\cert}{\mathsf{cert}}
\newcommand{\valid}{\mathsf{valid}}
\newcommand{\data}{\mathsf{data}}
\newcommand{\says}{\mathbin{\mathrm{says}}}
\DeclareMathOperator{\ok}{\mathit{ok}}
\DeclareMathOperator{\id}{\mathit{id}}
\DeclareMathOperator{\resource}{\mathit{resource}}
\DeclareMathOperator{\approved}{\mathit{approved}}
\DeclareMathOperator{\ask}{\mathit{ask}}
\DeclareMathOperator{\meas}{\mathit{meas}}
\DeclareMathOperator{\verifier}{\mathit{verifier}}

\DeclareMathOperator{\epca}{\mathit{epca}}
\DeclareMathOperator{\ltk}{\mathsf{ltk}}

\title{An Analysis of the CAVES Attestation\\ Protocol using CPSA}
\author{John D. Ramsdell\and Joshua D.~Guttman\and Jonathan K.~Millen\and
Brian O'Hanlon}

\begin{document}

\iffalse
\maketitle
\else




\begin{titlepage}
\begin{trivlist}\sffamily\bfseries\large
\item
MTR090213\\[-1.2ex]
\hrule ~\\
{\mdseries MITRE TECHNICAL REPORT}\\[1cm]
\LARGE
An Analysis of the CAVES Attestation\\[1ex] Protocol using CPSA\\[2.5cm]
\large
December 2009\\
~\\
\mdseries
John D.~Ramsdell\\
Joshua D.~Guttman\\
Jonathan K.~Millen\\
Brian O'Hanlon \\
\vfill
\normalsize
\bfseries
\begingroup\footnotesize
\begin{tabbing}
Sponsor: \phantom{spo} \= NSA/R23 \phantom{phantom} \=
Contract No.: \phantom{pro}\= W15P7T-08-C-F600 \\
Dept. No.: \>G020 \>Project No.: \>0708N6BZ\\[1.2cm]
The views, opinions and/or findings contained in \>\>\>\phantom{space}
\ifreleased
\= Approved for Public Release\\
this report are those of The MITRE Corporation\\
and should not be construed as an official\\
\fi
Government position, policy, or decision, unless\\
designated by other documentation.\\[4mm]
{\copyright} 2009 The MITRE Corporation. All Rights Reserved.
\end{tabbing}
\endgroup
~\\
\noindent
\includegraphics{mitrelogo-0.mps}\\
Center for Integrated Intelligence Systems\\
Bedford, Massachusetts
\end{trivlist}
\end{titlepage}

\noindent
Approved by:\\[1in]
\rule{3in}{.3mm}\\
Amy Herzog, 0707N6BZ Project Leader

\clearpage

\fi

\begin{abstract}
This paper describes the CAVES attestation protocol and presents a
tool-supported analysis showing that the runs of the protocol achieve
stated goals.  The goals are stated formally by annotating the
protocol with logical formulas using the rely-guarantee method.  The
protocol analysis tool used is the Cryptographic Protocol Shape
Analyzer.
\end{abstract}

\tableofcontents
\section{Introduction}

This paper describes the CAVES attestation protocol and presents a
tool-supported analysis showing that the runs of the protocol achieve
stated goals.  The protocol analysis tool used is the Cryptographic
Protocol Shape Analyzer (CPSA)~\cite{cpsa09}.

An attestation protocol is an exchange of messages over a network by
which an appraiser obtains evidence about the state of a target
platform.  The crucial principles for an attestation architecture,
according to \cite{CokerAtEl08}, are the following, paraphrased for
brevity:
\begin{description}
\item[Fresh information] Evidence should be up to date.
\item[Comprehensive information] Evidence should be collected by local
measurement tools with access to the entire internal state.
\item[Constrained disclosure] A target should be able to identify the
appraiser and restrict the evidence sent to it accordingly.
\item[Semantic explicitness] The target and type of the evidence should
be identified to the appraiser.
\item[Trustworthy mechanism] Attestation mechanisms should provide
evidence of their trustworthiness.
\end{description}
We shall see in the sections below how CPSA supports verification of
attestation protocol properties motivated by these principles. The
protocol is only part of the architecture that provides attestation,
but it contributes to all of the principles.

\subsection{CPSA}\label{sec:cpsa}

To analyze a cryptographic protocol, one finds out what security
properties---essentially, authentication and secrecy properties---are
true in all its possible executions. The {\em shapes} of a protocol,
relative to some assumptions, are the minimal, essentially different
executions compatible with those assumptions. A protocol may have one,
a few, many, or an infinite number of shapes relative to a choice of
assumptions.  Secrecy and authentication properties can be evaluated
by examining the shapes to see if they are all consistent with the
desired properties.  A ``problem statement'' means a set of
assumptions used to start off a search for shapes.  The ``scenarios''
of Section~\ref{sec:scenarios} explore the behavior of CAVES relative
to many different problem statements.

A central part of a problem statement is a designation of some
long-term keys that are assumed to be uncompromised.  These keys are
assumed to be used only in accordance with the protocol, i.e.~by a
principal that executes message transmissions and receptions in order
as stipulated by the protocol definition.  Another part of a problem
statement may be an assumption about some session-specific values,
such as session keys or nonces, asserting that these values are
created freshly, and will not be re-created independently, whether by
an adversary or else by an unlucky protocol participant in a collision
of randomly chosen values.  A problem statement also conveys some
behavior of uncompromised participants, using the uncompromised keys
and freshly generated values.

The CPSA program accepts a specification of the protocol and a set of
problem statements in an S-expression text format and, upon
termination, generates an XHTML document.  The output shows the shapes
derived for each problem statement. Most protocols and problems yield
just a few shapes. If there are very many, or an infinite number, of
shapes, CPSA will exit when specified storage bounds are exceeded.

\subsection{Rely-Guarantee Method}

Some application-specific protocol goals are stated formally by
annotating the protocol with logical formulas using the rely-guarantee
method~\cite{GHTCRS05}.  In CPSA, each role of a protocol specifies a
sequence of messaging events, each one being either a message
transmission or a reception.  The rely-guarantee method calls for
annotating each event with a formula.  The formulas are expressed in a
modal logic that allows principals to make local decisions based on
assertions made by peers.

The formula annotating a message transmission is a \emph{guarantee}
$\phi$.  The protocol instructs a principal $A$ to ascertain that
$\phi$ is true before sending the message.  The formula $\phi$ may
contain parameters that also occur in the message to be sent, and $A$
may fill in these parameters in a specific way to form a true instance
of $\phi$.  $A$ then transmits the corresponding instance of the
message.  If $A$ cannot ascertain a true instance of $\phi$, then $A$
must not continue this branch of the protocol.  If abrupt termination
should be avoided, the protocol may provide some error-recovery
branch.

The formula annotating a message reception is a \emph{rely} formula.
Typically, the rely formula is of the form $A\says\phi$,
or a conjunction of formulas of this form.  The rely formula may
contain parameters that also appear within the message to be received,
so that the message received on a particular occasion determines an
instance of the rely formula specified in the protocol definition.
The principal $B$ executing the message reception can add this
instance of the rely formula to its store of knowledge (its ``local
theory'').  Thus, this new fact may be used later to ascertain the
truth of guarantee formulas for future message transmissions.

A protocol is \emph{sound} relative to some assumptions if in every
execution compatible with those assumptions, each rely statement is
true.  For instance, if a particular message transmission $B$
relies on a formula $A\says\phi$, then there should have been some
earlier message transmissions in which the principal $A$ guaranteed
formulas $\phi'$, such that $\phi$ is a logical consequence of the
formulas $\phi'$.  That is, $A$ really is committed to the assertion
$\phi$.  The initial local theory of a participant is irrelevant
in determining if a protocol is sound.

We call a behavior \emph{regular} if it is an uncompromised local
execution that follows the protocol description.  We call a principal
\emph{regular} if all of its behaviors are regular, normally because
its long-term keys have been assumed uncompromised.  Only a regular
principal can be expected to ascertain the truth of formula before
sending a message.  The adversary will send messages even when the
expected guarantee is false, whenever this appears useful to achieving
its malicious goals.

The soundness property allows principals to make local decisions based
on guarantees made by other regular principals.  Given a set of shapes
computed by CPSA for an annotated protocol, another CPSA tool
instantiates the formulas; a protocol is sound relative to a set of
assumptions if all of the resulting formulas are logically valid.

\subsection{Literate Programming}

The style of literate programming combines source code and
documentation into a single source file.  This report is a literate
program. This means that in the course of reading this report, you
will read the full CPSA specification of the CAVES protocol, not just a
high-level description of it.  The literate programming metalanguage
provides a mechanism for presenting specifications or code to the
reader in a different order from the way it is supplied to the
analysis tool.  Thus the protocol can be described in a logical
manner.

A reader unfamiliar with CPSA may ignore the fragments of CPSA
specifications woven into this document and still understand the
conclusions of the paper.  A reader familiar with CPSA will see how
the tool was used to support the conclusions reached in this paper.

\section{CAVES Purpose and Features}

A simple protocol for attestation using a TPM was given by Sailer, et
al~\cite{Sailer04}.  Their protocol is sketched in
Figure~\ref{fig:sailer}, using terminology that is closer to our
conventions. A Challenger (which we call a Verifier) sends a challenge
containing a nonce~$N_V$ to a client's Attestation Service (which we
call an Attester). The Attester responds with a measurement list~$J_O$
and a TPM quote containing PCR values that authenticate the
measurements.  The quote binds the PCR vector~$P$ with a nonce using
hashing, and is signed with the identity key~$I$.  In this scheme one
expects that the measurements are taken once after each system boot,
so that the PCR values obtained by extending hash values into the PCRs
are predictable.

\begin{figure}
$$\xymatrix@C=10em{
\txt{\strut Attester}&\txt{\strut Verifier}\\
\bullet\ar@{=>}[d]&\bullet\ar@{=>}[d]\ar[l]_{\textstyle N_V}\\
\bullet\ar[r]^{\textstyle P,\enc{\#(N_V,P)}{I^{-1}},J_O}&\bullet}$$
\caption{Sailer Protocol}\label{fig:sailer}
\end{figure}

Confidentiality and authentication for this exchange are provided by
embedding it in an SSL session. The actual protocol specifies
additional details, such as which measurements and PCRs are requested
by the Server. The TNC (Trusted Network Connect) framework devised by
the TCG supports this type of attestation.

The CAVES protocol expands upon this basic scheme in several respects.
First, CAVES introduces a Server principal separate from the
Verifier. This allows several Servers to share a Verifier, so that
fewer Verifiers are needed to keep and maintain a database of
acceptable measurements. In effect, the Server is the enforcement
point and the Verifier is the decision point.

We have incorporated the certificate handling and session key
management into the CAVES protocol, rather than assuming that SSL
provides it. This approach gives us flexibility to manage more than
the original two principals.

Another feature of CAVES is a provision to request fresh measurements
during the attestation session. Depending on PCR usage, it may be
impractical to extend hashes of fresh measurements into PCRs, since
the result is not predictable unless the PCR can be reset first, or
the verifier is able to keep track of the entire history of
measurements since the last reboot.  In CAVES, these measurements are
authenticated cryptographically without using a PCR.

CAVES is intended to be used for a virtualized client architecture with
a hypervisor and multiple virtual machines (VMs). One advantage of
such an architecture is that a system-dependent or
application-dependent measurement agent can be placed in a trusted
Measurement VM, isolated from the Client VM containing the user
applications and OS.  The hypervisor can give the Measurement VM read
privileges necessary to examine the Client VM. The Measurement VM is
trusted to perform its measurement and bind the measurement result to
the TPM report in accordance with the protocol.  More generally, one
can have an Attestation Manager VM that obtains multiple measurements
of (different parts of) the same Client VM from various other
measurement agent VMs, and combines the results.

\section{The Protocol}\label{sec:protocol}

\begin{figure}
\newcommand{\syms}[3]{\ensuremath{#1}&\texttt{#2}&#3}
\begin{center}
\begin{tabular}{@{}ccc@{}}
\begin{tabular}{lll}
\syms{C}{c}{Client}\\
\syms{A}{a}{Attester}\\
\syms{V}{v}{Verifier}\\
\syms{E}{e}{EPCA}\\
\syms{S}{s}{Server}
\end{tabular}
&
\begin{tabular}{lll}
\syms{R}{r}{Request}\\
\syms{D}{d}{Data}\\
\syms{M}{m}{PCR Mask}\\
\syms{P}{p}{PCR Vector}\\
\syms{J}{j}{Query}\\
\syms{J_O}{jo}{Measurement}
\end{tabular}
&
\begin{tabular}{lll}
\syms{N_S}{ns}{Server Nonce}\\
\syms{N_V}{nv}{Verifier Nonce}\\
\syms{K}{k}{Client Key}\\
\syms{K'}{kp}{Attester Key}\\
\syms{I}{i}{Identity Key}\\
\syms{B}{b}{Blob}
\end{tabular}
\end{tabular}
\end{center}
\caption{CAVES Legend}\label{fig:legend}
\end{figure}

The overall plan of CAVES is shown in Figure~\ref{fig:caves}, and the
conventions used for variables are in Figure~\ref{fig:legend}. Its
roles are Client, Attester, Verifier, Enterprise Privacy Certificate
Authority (EPCA), and Server.  The name of the protocol was derived
from the first letter in each role name.  The EPCA role represents
the source of identity certificates.

\newcommand{\ctosa}{\enc{R,A,K}{K_S}}
\newcommand{\stova}{\enc{S,R,A,N_S}{K_V}}
\newcommand{\ptova}{\enc{\cert,A, I, E}{K^{-1}_E}}
\newcommand{\vtosa}{\enc{N_V,J,V,N_S,M}{K_S}}
\newcommand{\stoca}{\enc{N_V,J,V,M}{K}}
\newcommand{\ctoaa}{\enc{S, N_V,J,V,M,R}{\ltk(A,A)}}
\newcommand{\blob}{\enc{K',S,J_O,M,P,
\enc{\hash{\hash{A,V,R,N_V,J,J_O},M,P}}{I^{-1}}}{K_V}}
\newcommand{\atoca}{\enc{K',N_V,B}{\ltk(A,A)}}
\newcommand{\vtosb}{\enc{\valid,N_S,K'}{K_S}}
\newcommand{\stocb}{\enc{\data,D}{K'}}
\begin{figure}
$$\xymatrix@R=1.8ex{
\txt{\strut Attester}&\txt{\strut Client}&\txt{\strut Server}&\txt{\strut Verifier}&\txt{\strut EPCA}\\
&\bullet\ar@{=>}[dddd]\ar[r]^{\textstyle 1}&\bullet\ar@{=>}[d]&&\\
&&\bullet\ar@{=>}[dd]\ar[r]^{\textstyle 2}&\bullet\ar@{=>}[d]&\\
&&&\bullet\ar@{=>}[d]&\bullet\ar[l]_{\textstyle 3}\\
&&\bullet\ar@{=>}[d]&\bullet\ar@{=>}[ddddd]\ar[l]_{\textstyle 4}&\\
&\bullet\ar@{=>}[d]&\bullet\ar@{=>}[ddd]\ar[l]_{\textstyle 5}&&&\\
\bullet\ar@{=>}[d]&\bullet\ar@{=>}[d]\ar[l]_{\textstyle 6}&&&\\
\bullet\ar[r]^{\textstyle 7}&\bullet\ar@{=>}[d]&&&\\
&\bullet\ar@{=>}[ddd]\ar[r]^{\textstyle 8}&\bullet\ar@{=>}[d]&&\\
&&\bullet\ar@{=>}[d]\ar[r]^{\textstyle 9}&\bullet\ar@{=>}[d]&\\
&&\bullet\ar@{=>}[d]&\bullet\ar[l]_{\textstyle 10}&\\
&\bullet&\bullet\ar[l]_{\textstyle 11}&&}$$
\begin{eqnarray}
C\to S&:&\ctosa\\
S\to V&:&\stova\\
E\to V&:&\ptova\\
V\to S&:&\vtosa\\
S\to C&:&\stoca\\
C\to A&:&\ctoaa\label{eq:attester request}\\
A\to C&:&\atoca\label{eq:attester response}\\
C\to S&:&B\\
S\to V&:&B\\
V\to S&:&\vtosb\\
S\to C&:&\stocb
\end{eqnarray}
$$B=\blob$$
\caption{CAVES Protocol}\label{fig:caves}
\end{figure}

In this protocol, the keys~$K_S$ and~$K_V$ are public (asymmetric)
encryption keys, for which the participants may have PKI certificates.
We assume that participants use these PKI certificates to decide
whether to regard the matching private parts as uncompromised.  The
key~$K$ is a symmetric session key created by Client~$C$ for this
interaction with Server~$S$.  Key~$K'$ is a symmetric session key
created by Attester~$A$.  Nonces~$N_S$ and~$N_V$ are generated
randomly for this session by the server and the verifier.

%

Each message description includes a version of the message in CPSA's
S-expression syntax.  The message S-expression is given as a Nuweb
macro definition used to name a fragment or ``scrap'' of the final
specification.

\begin{description}
\item[$C\to S:\ctosa\colon$]
The protocol begins with a request by client~$C$ to receive some
resource~$R$, which includes a session key~$K$ to use for future
communication, e.g.\ to deliver the data~$D$ for $R$.  Distinguished
name~$A$ identifies the identity key to be used by the TPM while
gathering evidence about the client.
\begin{flushleft} \small
\begin{minipage}{\linewidth} \label{scrap1}
$\langle\,$initial request\nobreak\ {\footnotesize \NWtarget{nuweb6a}{6a}}$\,\rangle\equiv$
\vspace{-1ex}
\begin{list}{}{} \item
\mbox{}\verb@(enc r a k (pubk s))@{\NWsep}
\end{list}
\vspace{-1ex}
\footnotesize\addtolength{\baselineskip}{-1ex}
\begin{list}{}{\setlength{\itemsep}{-\parsep}\setlength{\itemindent}{-\leftmargin}}
\item \NWtxtMacroRefIn\ \NWlink{nuweb19a}{19a}\NWlink{nuweb20a}{, 20a}.
\end{list}
\end{minipage}\\[4ex]
\end{flushleft}
Nuweb generates the page numbers that follow phrase
``\NWtxtMacroRefIn''.  A use of a macro includes the page numbers
of the scraps that make up its definition.
\item[$S\to V:\stova\colon$]
Server~$S$ delegates to some acceptable verifier~$V$ the task of
appraising $C$.
\begin{flushleft} \small
\begin{minipage}{\linewidth} \label{scrap2}
$\langle\,$verification request\nobreak\ {\footnotesize \NWtarget{nuweb6b}{6b}}$\,\rangle\equiv$
\vspace{-1ex}
\begin{list}{}{} \item
\mbox{}\verb@(enc s r a ns (pubk v))@{\NWsep}
\end{list}
\vspace{-1ex}
\footnotesize\addtolength{\baselineskip}{-1ex}
\begin{list}{}{\setlength{\itemsep}{-\parsep}\setlength{\itemindent}{-\leftmargin}}
\item \NWtxtMacroRefIn\ \NWlink{nuweb20a}{20a}\NWlink{nuweb21a}{, 21a}.
\end{list}
\end{minipage}\\[4ex]
\end{flushleft}
\item[$E\to V:\ptova\colon$]
The verifier obtains the identity certificate with $A$ as its
distinguished name.  The identity key~$I$ is certified by an
Enterprise Privacy Certification Authority~$E$.  For simplicity, the
verifier's request for the certificate has been omitted.  Perhaps the
client sends the certificate with its request for the data.
\begin{flushleft} \small
\begin{minipage}{\linewidth} \label{scrap3}
$\langle\,$identity certificate\nobreak\ {\footnotesize \NWtarget{nuweb8a}{8a}}$\,\rangle\equiv$
\vspace{-1ex}
\begin{list}{}{} \item
\mbox{}\verb@(enc "cert" a i e (privk e))@{\NWsep}
\end{list}
\vspace{-1ex}
\footnotesize\addtolength{\baselineskip}{-1ex}
\begin{list}{}{\setlength{\itemsep}{-\parsep}\setlength{\itemindent}{-\leftmargin}}
\item \NWtxtMacroRefIn\ \NWlink{nuweb18}{18}\NWlink{nuweb21a}{, 21a}.
\end{list}
\end{minipage}\\[4ex]
\end{flushleft}
\item[$V\to S:\vtosa\colon$]
The verifier uses $A$ and information in $R$ to select an appropriate
query~$J$ and a selection mask~$M$, and delivers both to attester~$A$,
by forwarding the information through $S$ and $C$, to $C$'s local
measurement agent.
\begin{flushleft} \small
\begin{minipage}{\linewidth} \label{scrap4}
$\langle\,$verification query\nobreak\ {\footnotesize \NWtarget{nuweb8b}{8b}}$\,\rangle\equiv$
\vspace{-1ex}
\begin{list}{}{} \item
\mbox{}\verb@(enc nv j v ns m (pubk s))@{\NWsep}
\end{list}
\vspace{-1ex}
\footnotesize\addtolength{\baselineskip}{-1ex}
\begin{list}{}{\setlength{\itemsep}{-\parsep}\setlength{\itemindent}{-\leftmargin}}
\item \NWtxtMacroRefIn\ \NWlink{nuweb20a}{20a}\NWlink{nuweb21a}{, 21a}.
\end{list}
\end{minipage}\\[4ex]
\end{flushleft}
\item[$S\to C:\stoca$]
\begin{flushleft} \small
\begin{minipage}{\linewidth} \label{scrap5}
$\langle\,$server query\nobreak\ {\footnotesize \NWtarget{nuweb8c}{8c}}$\,\rangle\equiv$
\vspace{-1ex}
\begin{list}{}{} \item
\mbox{}\verb@(enc nv j v m k)@{\NWsep}
\end{list}
\vspace{-1ex}
\footnotesize\addtolength{\baselineskip}{-1ex}
\begin{list}{}{\setlength{\itemsep}{-\parsep}\setlength{\itemindent}{-\leftmargin}}
\item \NWtxtMacroRefIn\ \NWlink{nuweb19a}{19a}\NWlink{nuweb20a}{, 20a}.
\end{list}
\end{minipage}\\[4ex]
\end{flushleft}
\item[$C\to A:\ctoaa\colon$]
The key~${\ltk(A,A)}$ is not a real key at all.  It is an artifact of
our method for representing the private inter-VM channel between $C$
and its local attestation service~$A$.

\begin{flushleft} \small
\begin{minipage}{\linewidth} \label{scrap6}
$\langle\,$client query\nobreak\ {\footnotesize \NWtarget{nuweb8d}{8d}}$\,\rangle\equiv$
\vspace{-1ex}
\begin{list}{}{} \item
\mbox{}\verb@(enc s nv j v m r (ltk a a))@{\NWsep}
\end{list}
\vspace{-1ex}
\footnotesize\addtolength{\baselineskip}{-1ex}
\begin{list}{}{\setlength{\itemsep}{-\parsep}\setlength{\itemindent}{-\leftmargin}}
\item \NWtxtMacroRefIn\ \NWlink{nuweb19a}{19a}\NWlink{nuweb22a}{, 22a}.
\end{list}
\end{minipage}\\[4ex]
\end{flushleft}
\item[$A\to C:\atoca\colon$]
Attester $A$ retrieves the evidence requested in the form of
a large term $B$ which we refer to as the ``blob''. As shown
in Fig.~\ref{fig:caves}, the blob is defined by
$$B = \blob.$$ This message packages $J$'s output~$J_O$ together with
the current PCR values~$P$ for the registers selected by $V$ in the
PCR mask~$M$.  This information is generated and authenticated using a
TPM quote.  The TPM quote uses the hash $\hash{A,V,R,N_V,J,J_O}$ of some
parameters as a nonce-like seed to be included in a digital signature.
The digital signature is prepared using $I^{-1}$, a TPM-resident
Attestation Identity private key.

CPSA has no explicit support for hashing, so a
hash is encoded as an asymmetric encryption in which no participant
has access to the decryption key.  The public key used to create a
hash is \texttt{hash} and a tag is added to ensure the hash is not
confused with other encrypted messages.  Thus the hash of
variable~$X$,  $\hash{X},$ is encoded as \texttt{(enc "hash" x hash)}.

\begin{flushleft} \small
\begin{minipage}{\linewidth} \label{scrap7}
$\langle\,$encoded report\nobreak\ {\footnotesize \NWtarget{nuweb9a}{9a}}$\,\rangle\equiv$
\vspace{-1ex}
\begin{list}{}{} \item
\mbox{}\verb@(enc kp nv @\hbox{$\langle\,$report\nobreak\ {\footnotesize \NWlink{nuweb9b}{9b}}$\,\rangle$}\verb@ (ltk a a))@{\NWsep}
\end{list}
\vspace{-1ex}
\footnotesize\addtolength{\baselineskip}{-1ex}
\begin{list}{}{\setlength{\itemsep}{-\parsep}\setlength{\itemindent}{-\leftmargin}}
\item \NWtxtMacroRefIn\ \NWlink{nuweb22a}{22a}.
\end{list}
\end{minipage}\\[4ex]
\end{flushleft}
\begin{flushleft} \small
\begin{minipage}{\linewidth} \label{scrap8}
$\langle\,$report\nobreak\ {\footnotesize \NWtarget{nuweb9b}{9b}}$\,\rangle\equiv$
\vspace{-1ex}
\begin{list}{}{} \item
\mbox{}\verb@@\\
\mbox{}\verb@(enc kp s jo m p@\\
\mbox{}\verb@  (enc@\\
\mbox{}\verb@     (enc "hash"@\\
\mbox{}\verb@        (enc "hash" a v r nv j jo hash)@\\
\mbox{}\verb@         m p hash)@\\
\mbox{}\verb@    (invk i))@\\
\mbox{}\verb@  (pubk v))@{\NWsep}
\end{list}
\vspace{-1ex}
\footnotesize\addtolength{\baselineskip}{-1ex}
\begin{list}{}{\setlength{\itemsep}{-\parsep}\setlength{\itemindent}{-\leftmargin}}
\item \NWtxtMacroRefIn\ \NWlink{nuweb9a}{9a}\NWlink{nuweb21a}{, 21a}.
\end{list}
\end{minipage}\\[4ex]
\end{flushleft}
Received message:
\begin{flushleft} \small
\begin{minipage}{\linewidth} \label{scrap9}
$\langle\,$received report\nobreak\ {\footnotesize \NWtarget{nuweb9c}{9c}}$\,\rangle\equiv$
\vspace{-1ex}
\begin{list}{}{} \item
\mbox{}\verb@(enc kp nv @\hbox{$\langle\,$blob\nobreak\ {\footnotesize \NWlink{nuweb10a}{10a}}$\,\rangle$}\verb@ (ltk a a))@{\NWsep}
\end{list}
\vspace{-1ex}
\footnotesize\addtolength{\baselineskip}{-1ex}
\begin{list}{}{\setlength{\itemsep}{-\parsep}\setlength{\itemindent}{-\leftmargin}}
\item \NWtxtMacroRefIn\ \NWlink{nuweb19a}{19a}.
\end{list}
\end{minipage}\\[4ex]
\end{flushleft}
\begin{flushleft} \small
\begin{minipage}{\linewidth} \label{scrap10}
$\langle\,$blob\nobreak\ {\footnotesize \NWtarget{nuweb10a}{10a}}$\,\rangle\equiv$
\vspace{-1ex}
\begin{list}{}{} \item
\mbox{}\verb@b@{\NWsep}
\end{list}
\vspace{-1ex}
\footnotesize\addtolength{\baselineskip}{-1ex}
\begin{list}{}{\setlength{\itemsep}{-\parsep}\setlength{\itemindent}{-\leftmargin}}
\item \NWtxtMacroRefIn\ \NWlink{nuweb9c}{9c}\NWlink{nuweb19a}{, 19a}\NWlink{nuweb20a}{, 20a}.
\end{list}
\end{minipage}\\[4ex]
\end{flushleft}
\item[$C\to S:B$]
\item[$S\to V:B\colon$]
When the blob is received by $C$ and forwarded through $S$,
$C$ and $S$ treat it as atom $B$ because it is encrypted with $V$'s
public encryption key, to ensure that $C$ and $S$ cannot read it.  $V$
uses the identity key in the Attestation Identity Certificate to
validate the quote.
\item[$V\to S:\vtosb\colon$]
If the evidence is valid, the server is notified of this fact.
\begin{flushleft} \small
\begin{minipage}{\linewidth} \label{scrap11}
$\langle\,$verification acceptance\nobreak\ {\footnotesize \NWtarget{nuweb10b}{10b}}$\,\rangle\equiv$
\vspace{-1ex}
\begin{list}{}{} \item
\mbox{}\verb@(enc "valid" kp ns (pubk s))@{\NWsep}
\end{list}
\vspace{-1ex}
\footnotesize\addtolength{\baselineskip}{-1ex}
\begin{list}{}{\setlength{\itemsep}{-\parsep}\setlength{\itemindent}{-\leftmargin}}
\item \NWtxtMacroRefIn\ \NWlink{nuweb20a}{20a}\NWlink{nuweb21a}{, 21a}.
\end{list}
\end{minipage}\\[4ex]
\end{flushleft}
\item[$S\to C:\stocb\colon$]
The server releases the data to a validated client.
\begin{flushleft} \small
\begin{minipage}{\linewidth} \label{scrap12}
$\langle\,$server response\nobreak\ {\footnotesize \NWtarget{nuweb10c}{10c}}$\,\rangle\equiv$
\vspace{-1ex}
\begin{list}{}{} \item
\mbox{}\verb@(enc "data" d kp)@{\NWsep}
\end{list}
\vspace{-1ex}
\footnotesize\addtolength{\baselineskip}{-1ex}
\begin{list}{}{\setlength{\itemsep}{-\parsep}\setlength{\itemindent}{-\leftmargin}}
\item \NWtxtMacroRefIn\ \NWlink{nuweb19a}{19a}\NWlink{nuweb20a}{, 20a}.
\end{list}
\end{minipage}\\[4ex]
\end{flushleft}
\end{description}

\section{CPSA Overview}
An introduction or primer to CPSA is delivered with the
program~\cite{RamsdellGuttman09}.  A brief introduction is included
here, along with a description of CPSA's support for the
rely-guarantee method.  The message terms \nterm{term} used by CPSA
are a straightforward representation of terms using Lisp-style, prefix
notation.

A subset of the terms are called {\em atoms}. Atoms belong to the {\em
  base sorts} \texttt{name, text, data, skey, akey}.  Syntactically,
atomic terms may be either symbols (i.e., identifiers) or
atomic-sorted function applications such as \texttt{(pubk~a)}.  Even
though an atom as a term may have terms within it, a receiver of an
atom is not allowed to extract terms that occur in it.  This reflects
the fact that the reception of the atom \texttt{(invk~k)}, the inverse
of some asymmetric key~\texttt{k}, does not allow the receiver to
construct~\texttt{k}.

Non-atomic terms are constructed by applications of encryption
(\texttt{enc}) and pairing (\texttt{cat}), where $n$-ary concatenation
is parsed right-associatively.  The second argument of an encryption
is the key. Encryption may also be written in an $n$-ary form where
the last argument is the key and the arguments preceding it are
implicitly concatenated.

A term carries one of its subterms if the possession of the right set
of keys allows the extraction of the subterm.  The carries relation is
the least relation such that (1)~$t$ carries~$t$, (2)~\texttt{(enc
  $t_0$ $t_1$)} carries~$t$ if~$t_0$ carries~$t$, and (3)~\texttt{(cat
  $t_0$ $t_1$)} carries~$t$ if~$t_0$ or~$t_1$ carries~$t$.  Note that
\texttt{(enc $t_0$ $t_1$)} does not carry~$t_1$ unless (anomalously)~$t_0$
carries~$t_1$.

\subsection{Protocols}

A protocol is a set of roles.
\begin{quote}
\begin{alltt}
(defprotocol \nterm{sym} basic \ntermp{role})
\end{alltt}
\end{quote}
The symbol \nterm{sym} names the protocol.  The symbol \texttt{basic}
identifies the term algebra used to specify messages in roles.

A role has the form:
\begin{quote}
\begin{alltt}
(defrole \nterm{sym} (vars \nterms{decl})
\quad (trace \ntermp{event})
\quad \ntermo{(non-orig \nterms{non})}
\quad \ntermo{(uniq-orig \nterms{atom})}
\quad \nterm{annos})

\nterm{non} ::= \nterm{atom} | (\nterm{height} \nterm{atom})
\end{alltt}
\end{quote}
Non-terminal \nterm{sym} is an S-expression symbol that names the
role.  A \nterm{decl} is a list of variable symbols followed by a sort
symbol.  The \texttt{trace} is a sequence of message events, each
indicating a message to be transmitted or received.  The syntax used
for a message event \nterm{event} has
one of two forms, \texttt{(send \nterm{term})} or \texttt{(recv
  \nterm{term})}.  The length of a role is the length of its trace,
and must be positive.  The remaining components of a role will be
described later.

A term originates in a trace if it is carried in some event
and the first event in which it is carried is a sending term.
A term is acquired by a trace if it first occurs in a receiving term
and is also carried by that term.

\subsection{Executions}

An execution of a protocol may involve any number of strands, each
conveying either regular or adversarial behavior.  Thus, each strand
is an instance of some role.  For CPSA input and output, a strand is
specified by the following form:
\begin{quote}
\begin{alltt}
(defstrand \nterm{sym} \nterm{int} \nterms{maplet})
\end{alltt}
\end{quote}
The symbol names the role, \nterm{int} is the height which must be
positive and no greater than the role's length, and the remainder
determines a substitution from role variables to terms.
\begin{quote}
\begin{alltt}
\nterm{maplet} ::= (\nterm{sym} \nterm{term})
\end{alltt}
\end{quote}
The trace associated with the specified behavior is the result of
truncating the role's trace so it agrees with the height, and applying
the substitution \texttt{(\nterms{maplet})}.

A strand's behavior includes inherited origination assumptions.  When
a role assumes atom~$a$ is uniquely originating using the
\texttt{uniq-orig} form, applying the substitution
\texttt{(\nterms{maplet})} to $a$ produces an inherited uniquely
originating atom.  Role atoms assumed to be non-originating using the
\texttt{non-orig} form are inherited similarly.  For a non-originating
assumption, a strand height may be associated with an atom.  In this
case, a non-originating assumption is inherited by strands that meet
or exceed the height constraint.  Note that the definition of a
uniquely originating atom and a non-originating atom in an execution
is still to come.

A strand in an execution is identified by a natural number.  To
describe an execution, the behavior of each participant is listed
sequentially, and position of the \texttt{defstrand} form in the list
determines the strand's identifier.  Zero-based indexing is used,
so zero identifies the first strand.

A messaging event in an execution occurs at a node, which is a pair of
natural numbers.  The first number is the strand's identifier.  The
second number is the position of an event in the trace of the
strand, once again using zero-based indexing.  Thus node \texttt{(1~1)}
in
\begin{quote}
\begin{alltt}
(defstrand r1 3 (a b) (b a))
(defstrand r2 2 (x a) (y a) (z b))
\end{alltt}
\end{quote}
names the last event in the last strand.  The term is the
result of instantiating the second event in role \texttt{r2}'s
trace using the substitution \texttt{((x a) (y a) (z b))}.

Message exchanges are part of an execution.  Each exchange is
described by a pair of nodes.  The first node must name a sending
term, and the second node must name a receiving term.  In an
execution, the two terms are the same.  Furthermore, for each
receiving term in a strand's trace, there is a unique node that
transmits its term.  In other words, all message receptions are
explained by transmissions within the execution.

In an execution, a \emph{uniquely originating atom} originates in the
trace of exactly one strand.  An inherited uniquely originating atom
must originate in the trace of its strand.  CPSA uses uniquely
originating atoms to model freshly generated nonces used in many
protocols.

A \emph{non-originating atom} is carried by no trace of any strand in
an execution, and it or its inverse is the key of an encryption in one
of those traces.  The inherited non-origination atoms must satisfy
this property too.

Strands in executions represent both adversarial and non-adversarial
behaviors.  A strand that is an instance of a protocol role is
non-adversarial, and is called regular.  A strand that represents
adversarial behavior is called a penetrator strand.

The roles that define adversary behavior codify the basic abilities
that make up the Dolev-Yao model.  They include transmitting an atom
such as a name or a key; transmitting a tag; transmitting an encrypted
message after receiving its plain text and the key; and transmitting a
plain text after receiving ciphertext and its decryption key.  The
adversary can also concatenate two messages, or separate the pieces of
a concatenated message.  Since a penetrator strand that encrypts or
decrypts must receive the key as one of its inputs, keys used by the
adversary---compromised keys---have always been transmitted by some
strand. The basic adversary roles are built into CPSA.

\subsection{Skeletons}

CPSA never directly represents adversarial behavior.  Instead, a
skeleton is used.  A skeleton represents regular behavior that might
make up part of an execution. A skeleton is specified in CPSA output using a
\texttt{defskeleton} form.
\begin{quote}
\begin{alltt}
(defskeleton \nterm{sym} (vars \nterms{decl})
\quad \ntermp{defstrand}
\quad \ntermo{(precedes \nterms{pair})}
\quad \ntermo{(non-orig \nterms{atom})}
\quad \ntermo{(uniq-orig \nterms{atom})})
\end{alltt}
\end{quote}
The symbol names the protocol used by its participants.  The regular
strands are specified as they are in an execution.  The precedes form
defines a binary relation on nodes (\texttt{\nterm{pair} ::=
  (\nterm{node} \nterm{node})}).  As in an execution, the first node
names a sending term and the second term names a receiving term.
Unlike an execution, the pair of nodes in the relation need not agree
on their message term.  Two nodes are related if the sending event
precedes the reception reception event, as an execution it represents
may include events in between.

The final two additional components of a skeleton are a set of
non-originating atoms, and a set of uniquely originating atoms.  To be
a skeleton, each uniquely originating atom must originate in at
most one strand in the skeleton, and each non-originating atom must
never be carried by some event in the skeleton and every
variable that occurs in the atom must occur in some event.
Furthermore, for each uniquely originating atom that originates in the
skeleton, the node relation must ensure that reception nodes that
carry the atom follow the node of its origination.

One special skeleton is associated with each execution.  It summarizes
the regular behavior of the execution.  It is derived from the
execution by enriching its node relation to contain all node orderings
implied by transitive closure, deleting all strands and nodes that
refer to penetrator behavior, and then performing the transitive
reduction on the resulting node relation.  The set of uniquely
originating atoms is the set of terms that originate on exactly one
strand in the execution, and are carried in a term of a regular
strand.  The set of non-originating atoms is the union of two sets.
One set contains each term that is used as an encryption or decryption
key in some term in the execution, but is not carried by any term.
The other set contains the terms specified by non-origination
assumptions in roles.  If a realized skeleton instance maps all of the
variables that occur in one of its non-originating role terms, the
mapped term is a member of the skeleton's set of non-originating
terms.  A skeleton is \emph{realized} if it summarizes the behavior of
some execution.

\subsubsection{Preskeletons}

Preskeletons are used to pose problems for CPSA to solve.  A
preskeleton is similar to a skeleton except atoms assumed to uniquely
originate may originate in more than one strand, and the node relation
need not ensure that reception nodes that carry the atom follow some
node of origination.  Experience has shown that it is more natural to
specify some problems in a form that doesn't satisfy all the
properties of a skeleton.  If CPSA cannot immediately convert its
input into a skeleton, an error is signal.  With the exception of the
restatement of the original problem, all preskeletons printed by CPSA
are skeletons.  A problem statement is called a \emph{scenario}, and
the converted skeleton is called the \emph{scenario skeleton}.

\subsubsection{Shapes}

Given a scenario skeleton, CPSA determines whether there is an execution
containing the strands in the skeleton, and satisfying its
origination assumptions.  Usually an execution contains additional
regular strands, as well as adversary behavior.  A major part of what
CPSA does is to find all additional regular strands that are necessary
to extend the scenario to an execution---a realized skeleton.  If a
realized skeleton is most-general, in the sense that there is no other
realized skeleton that can be instantiated to it by merging nodes or
atoms, then it is called a {\em shape}. CPSA finds all shapes for a
scenario.

\subsection{Listeners}

In addition to the roles specified in a protocol, for each term~$t$, a
regular strand may be an instance of a so-called {\em listener} role
with the trace \texttt{(recv~$t$) (send~$t$)}.  There are no
non-originating or uniquely originating atoms associated with a
listener role.  Listener behavior is specified with:
\begin{quote}
\begin{alltt}
(deflistener \nterm{term})
\end{alltt}
\end{quote}

A listener strand is used in a skeleton to assert that a term~$t$ is
derivable by the adversary, unprotected by encryption.  Hence it is used
to test for compromise of a term.  The term is protected if the
resulting skeleton is unrealizable.  Otherwise, CPSA will find a shape
that shows how the adversary accesses~$t$.

\subsection{Annotations}

To be analyzed, each role in a protocol must include an
\texttt{annotations} form, as defined in Table~\ref{tab:anno}.  The
\nterm{term} just after the \texttt{annotations} symbol is a role atom
that, when instantiated, is the principal associated with the strand in
the shape.  A principal may be a key.

\begin{table}
\begingroup\ttfamily
\begin{tabular}{rcl}
\nterm{annos}&$::=$
&(annotations \nterm{term} (\nterm{int} \nterm{form})$^\ast$)
\\ \nterm{form}&$::=$&(\nterm{sym} \nterms{fterm}) | (not \nterm{form})
\\ &|& (and \nterms{form}) | (or \nterms{form})
\\ &|& (implies \nterms{form} \nterm{form})
\\ &|& (iff \nterm{form} \nterm{form})
\\ &|& (says \nterm{term} \nterm{form})
\\ &|& (forall (\nterms{decl}) \nterm{form})
\\ &|& (exists (\nterms{decl}) \nterm{form})
\\ \nterm{fterm}&$::=$&\nterm{term} | (\nterm{sym} \nterms{fterm})
\end{tabular}
\endgroup
\caption{Annotation Syntax}\label{tab:anno}
\end{table}

What follows is sequences of pairs.  The integer gives the position of
the event in the trace that is annotated by the formula, using
zero-based indexing.  Thus, each formula is associated with a node.
Nodes for which no formula is specified are implicitly defined to be
the trivial formula \texttt{(and)} for truth.  Use \texttt{(or)} for
falsehood.

The language of formulas is first-order logic extended with a modal
``says'' operator. Formula terms may include function symbols that are not
part of a protocol's message signature.

On output, each shape contains an \texttt{annotations} form and an
\texttt{obligations} form. The annotations form presents every
non-trivial formula derived from the protocol. The obligations form
presents every non-trivial formula that must be true if the shape is
sound.

In what follows, annotated roles will be presented in two forms: a
tabular form and a CPSA S-expression form.  A template for the
tabular form follows.

\setcounter{role}{-1}
\begin{role}[Name $P$]
\begin{eqnarray*}
&\outbnd M_0&\Phi_0\\
&\inbnd M_1&\Phi_1\\
&\outbnd M_2&\Phi_2
\end{eqnarray*}
\end{role}

A plus sign denotes a sent term, and a minus sign denotes a received
term.  The S-expression version of the role follows.

\begin{quote}
\newcommand{\trace}{\texttt{(send $M_0$) (recv $M_1$) (send $M_2$)}}
\newcommand{\annos}{\texttt{$P$ (0 $\Phi_0$) (1 $\Phi_1$) (2 $\Phi_2$)}}
\begin{alltt}
(defrole name (vars \textrm{not specified})
\quad (trace \trace)
\quad (annotations \annos))
\end{alltt}
\end{quote}

\subsection{Run Time Options}

CPSA run time options are specified with a herald form.  For this
analysis, the bound on the number of strands considered by CPSA has
been raised to twelve, and CPSA has been directed to preferentially
focus on nonces used by the protocol as opposed to encryptions.

\begin{flushleft} \small
\begin{minipage}{\linewidth} \label{scrap13}
\verb@"caves.scm"@\nobreak\ {\footnotesize \NWtarget{nuweb17}{17} }$\equiv$
\vspace{-1ex}
\begin{list}{}{} \item
\mbox{}\verb@(herald "CAVES Attestation Protocol"@\\
\mbox{}\verb@  (bound 12) (check-nonces))@{\NWsep}
\end{list}
\vspace{-1ex}
\footnotesize\addtolength{\baselineskip}{-1ex}
\begin{list}{}{\setlength{\itemsep}{-\parsep}\setlength{\itemindent}{-\leftmargin}}
\item \NWtxtFileDefBy\ \NWlink{nuweb17}{17}\NWlink{nuweb23}{, 23}\NWlink{nuweb24}{, 24}\NWlink{nuweb25a}{, 25a}\NWlink{nuweb25b}{b}\NWlink{nuweb25c}{c}\NWlink{nuweb26a}{, 26a}\NWlink{nuweb26b}{b}\NWlink{nuweb26c}{c}\NWlink{nuweb27a}{, 27a}\NWlink{nuweb27b}{b}.
\end{list}
\end{minipage}\\[4ex]
\end{flushleft}
\section{Annotated Roles}\label{sec:roles}

This section describes the protocol from the perspective of each
individual role.  The description of each role lists message events
and origination assumptions.  In support of the rely-guarantee method,
a role lists logical formulas that annotate each node.
The rely-guarantee annotations will be used in
Section~\ref{sec:trust argument}, and can be ignored on a first pass.

The first role presented is the enterprise privacy certificate
authority, due to its simplicity.  It transmits the signed attestation
identity certificate for distinguished name~$A$, and it
guarantees $\id(A,I)$, i.e.\ that~$A$'s identity key is~$I$.

\begin{role}[EPCA $E$]
\begin{eqnarray}
&\outbnd\ptova&\id(A,I)\label{eq:epca 1}
\end{eqnarray}
where~$I^{-1}$ is uncompromised.  The initial theory of the EPCA
defines the $\id$ relation.
\end{role}

\begin{flushleft} \small
\begin{minipage}{\linewidth} \label{scrap14}
$\langle\,$enterprise privacy certificate authority role\nobreak\ {\footnotesize \NWtarget{nuweb18}{18}}$\,\rangle\equiv$
\vspace{-1ex}
\begin{list}{}{} \item
\mbox{}\verb@(defrole epca (vars (a e name) (i akey))@\\
\mbox{}\verb@  (trace@\\
\mbox{}\verb@    (send @\hbox{$\langle\,$identity certificate\nobreak\ {\footnotesize \NWlink{nuweb8a}{8a}}$\,\rangle$}\verb@))@\\
\mbox{}\verb@  (non-orig (invk i))@\\
\mbox{}\verb@  (annotations e@\\
\mbox{}\verb@    (0 (id a i))))@{\NWsep}
\end{list}
\vspace{-1ex}
\footnotesize\addtolength{\baselineskip}{-1ex}
\begin{list}{}{\setlength{\itemsep}{-\parsep}\setlength{\itemindent}{-\leftmargin}}
\item \NWtxtMacroRefIn\ \NWlink{nuweb23}{23}.
\end{list}
\end{minipage}\\[4ex]
\end{flushleft}
The client requests data in Message~\ref{eq:client 0} and receives the
data in Message~\ref{eq:client 5}.  The other messages relay information
between the server and the attester.  In this protocol, no participant
takes note of the name associated the client, the variable~$C$.  The
variable~$C$ appears in Message~\ref{eq:client 0} as an artifact of our
implementation---the rely-guarantee analysis software requires that
every role declare a principal.  The sole purpose for transmitting~$C$
is to satisfy this requirement.  Because no other role makes use
of~$C$, its presence does not affect the shapes produced by CPSA.

\begin{role}[Client $C$]
\begin{eqnarray}
&\outbnd(C,\ctosa)\label{eq:client 0}\\
&\inbnd\stoca\\
&\outbnd\ctoaa\\
&\inbnd\atoca\\
&\outbnd B\\
&\inbnd\stocb&S\says\resource(R,D)\label{eq:client 5}
\end{eqnarray}
where~$K$ is fresh.
\end{role}

\begin{flushleft} \small
\begin{minipage}{\linewidth} \label{scrap15}
$\langle\,$client role\nobreak\ {\footnotesize \NWtarget{nuweb19a}{19a}}$\,\rangle\equiv$
\vspace{-1ex}
\begin{list}{}{} \item
\mbox{}\verb@(defrole client (vars @\hbox{$\langle\,$client declarations\nobreak\ {\footnotesize \NWlink{nuweb19b}{19b}}$\,\rangle$}\verb@)@\\
\mbox{}\verb@ (trace@\\
\mbox{}\verb@   (send (cat c @\hbox{$\langle\,$initial request\nobreak\ {\footnotesize \NWlink{nuweb6a}{6a}}$\,\rangle$}\verb@))@\\
\mbox{}\verb@   (recv @\hbox{$\langle\,$server query\nobreak\ {\footnotesize \NWlink{nuweb8c}{8c}}$\,\rangle$}\verb@)@\\
\mbox{}\verb@   (send @\hbox{$\langle\,$client query\nobreak\ {\footnotesize \NWlink{nuweb8d}{8d}}$\,\rangle$}\verb@)@\\
\mbox{}\verb@   (recv @\hbox{$\langle\,$received report\nobreak\ {\footnotesize \NWlink{nuweb9c}{9c}}$\,\rangle$}\verb@)@\\
\mbox{}\verb@   (send @\hbox{$\langle\,$blob\nobreak\ {\footnotesize \NWlink{nuweb10a}{10a}}$\,\rangle$}\verb@)@\\
\mbox{}\verb@   (recv @\hbox{$\langle\,$server response\nobreak\ {\footnotesize \NWlink{nuweb10c}{10c}}$\,\rangle$}\verb@))@\\
\mbox{}\verb@ (uniq-orig k)@\\
\mbox{}\verb@ (annotations c@\\
\mbox{}\verb@   (5 (says s (resource r d)))))@{\NWsep}
\end{list}
\vspace{-1ex}
\footnotesize\addtolength{\baselineskip}{-1ex}
\begin{list}{}{\setlength{\itemsep}{-\parsep}\setlength{\itemindent}{-\leftmargin}}
\item \NWtxtMacroRefIn\ \NWlink{nuweb23}{23}.
\end{list}
\end{minipage}\\[4ex]
\end{flushleft}
\begin{flushleft} \small
\begin{minipage}{\linewidth} \label{scrap16}
$\langle\,$client declarations\nobreak\ {\footnotesize \NWtarget{nuweb19b}{19b}}$\,\rangle\equiv$
\vspace{-1ex}
\begin{list}{}{} \item
\mbox{}\verb@(c a v s name) (r m j d text)@\\
\mbox{}\verb@(nv data) (k kp skey) (b mesg)@{\NWsep}
\end{list}
\vspace{-1ex}
\footnotesize\addtolength{\baselineskip}{-1ex}
\begin{list}{}{\setlength{\itemsep}{-\parsep}\setlength{\itemindent}{-\leftmargin}}
\item \NWtxtMacroRefIn\ \NWlink{nuweb19a}{19a}.
\end{list}
\end{minipage}\\[4ex]
\end{flushleft}
\begin{role}[Server $S$]
\begin{eqnarray}
&\inbnd\ctosa\\
&\outbnd\stova&\verifier(V)\\
&\inbnd\vtosa\\
&\outbnd\stoca\\
&\inbnd B\\
&\outbnd B\\
&\inbnd\vtosb&V\says\approved(R,A,N_V)\\
&\outbnd\stocb&\approved(R,A,N_V)\land\resource(R,D)
\end{eqnarray}
where~$N_S$ is fresh.  The initial
theory of the server defines the $\verifier$ and $\resource$
relations and contains the following rule.
\begin{eqnarray}
&\approved(r,a,n)\gets\verifier(v)\land
  v\says\approved(r,a,n)\label{eq:trust verifier}
\end{eqnarray}
\end{role}

\begin{flushleft} \small
\begin{minipage}{\linewidth} \label{scrap17}
$\langle\,$server role\nobreak\ {\footnotesize \NWtarget{nuweb20a}{20a}}$\,\rangle\equiv$
\vspace{-1ex}
\begin{list}{}{} \item
\mbox{}\verb@(defrole server (vars @\hbox{$\langle\,$server declarations\nobreak\ {\footnotesize \NWlink{nuweb20b}{20b}}$\,\rangle$}\verb@)@\\
\mbox{}\verb@  (trace@\\
\mbox{}\verb@    (recv @\hbox{$\langle\,$initial request\nobreak\ {\footnotesize \NWlink{nuweb6a}{6a}}$\,\rangle$}\verb@)@\\
\mbox{}\verb@    (send @\hbox{$\langle\,$verification request\nobreak\ {\footnotesize \NWlink{nuweb6b}{6b}}$\,\rangle$}\verb@)@\\
\mbox{}\verb@    (recv @\hbox{$\langle\,$verification query\nobreak\ {\footnotesize \NWlink{nuweb8b}{8b}}$\,\rangle$}\verb@)@\\
\mbox{}\verb@    (send @\hbox{$\langle\,$server query\nobreak\ {\footnotesize \NWlink{nuweb8c}{8c}}$\,\rangle$}\verb@)@\\
\mbox{}\verb@    (recv @\hbox{$\langle\,$blob\nobreak\ {\footnotesize \NWlink{nuweb10a}{10a}}$\,\rangle$}\verb@)@\\
\mbox{}\verb@    (send @\hbox{$\langle\,$blob\nobreak\ {\footnotesize \NWlink{nuweb10a}{10a}}$\,\rangle$}\verb@)@\\
\mbox{}\verb@    (recv @\hbox{$\langle\,$verification acceptance\nobreak\ {\footnotesize \NWlink{nuweb10b}{10b}}$\,\rangle$}\verb@)@\\
\mbox{}\verb@    (send @\hbox{$\langle\,$server response\nobreak\ {\footnotesize \NWlink{nuweb10c}{10c}}$\,\rangle$}\verb@))@\\
\mbox{}\verb@    (uniq-orig ns)@\\
\mbox{}\verb@    (annotations s@\\
\mbox{}\verb@      (1 (verifier v))@\\
\mbox{}\verb@      (6 (says v (approved r a nv)))@\\
\mbox{}\verb@      (7 (and (approved r a nv) (resource r d)))))@{\NWsep}
\end{list}
\vspace{-1ex}
\footnotesize\addtolength{\baselineskip}{-1ex}
\begin{list}{}{\setlength{\itemsep}{-\parsep}\setlength{\itemindent}{-\leftmargin}}
\item \NWtxtMacroRefIn\ \NWlink{nuweb23}{23}.
\end{list}
\end{minipage}\\[4ex]
\end{flushleft}
\begin{flushleft} \small
\begin{minipage}{\linewidth} \label{scrap18}
$\langle\,$server declarations\nobreak\ {\footnotesize \NWtarget{nuweb20b}{20b}}$\,\rangle\equiv$
\vspace{-1ex}
\begin{list}{}{} \item
\mbox{}\verb@(a v s name) (r m j d text)@\\
\mbox{}\verb@(ns nv data) (k kp skey) (b mesg)@{\NWsep}
\end{list}
\vspace{-1ex}
\footnotesize\addtolength{\baselineskip}{-1ex}
\begin{list}{}{\setlength{\itemsep}{-\parsep}\setlength{\itemindent}{-\leftmargin}}
\item \NWtxtMacroRefIn\ \NWlink{nuweb20a}{20a}.
\end{list}
\end{minipage}\\[4ex]
\end{flushleft}
\begin{role}[Verifier $V$]
\begin{eqnarray}
&\inbnd\stova\\
&\inbnd\ptova&E\says\id(A,I)\\
&\outbnd\vtosa&\ask(R,A,J,M)\\
&\inbnd B&A\says\meas(I,N_V,J,J_O,M,P)\\
&\outbnd\vtosb&\approved(R,A,N_V)\label{eq:verifier 4}
\end{eqnarray}
where $B=\blob$, $N_V$ is fresh and~$K^{-1}_P$ is uncompromised.  The initial
theory of the verifier defines the $\epca$ and $\ok$ relations, and
contains the following rules.
\begin{eqnarray}
&&\ask(r,a,j,m)\gets\ok(r,a,j,j_o,m,p)\label{eq:ask}\\
&&\id(a,i)\gets\epca(e)\land e\says\id(a,i)\label{eq:trust epca}\\
&&\approved(r,a,n)\gets\id(a,i)\land{} \label{eq:approved rule}\\
&&\qquad a\says\meas(i,n,j,j_o,m,p)\land\ok(r,a,j,j_o,m,p) \nonumber
\end{eqnarray}
\end{role}

\begin{flushleft} \small
\begin{minipage}{\linewidth} \label{scrap19}
$\langle\,$verifier role\nobreak\ {\footnotesize \NWtarget{nuweb21a}{21a}}$\,\rangle\equiv$
\vspace{-1ex}
\begin{list}{}{} \item
\mbox{}\verb@(defrole verifier (vars @\hbox{$\langle\,$verifier declarations\nobreak\ {\footnotesize \NWlink{nuweb21b}{21b}}$\,\rangle$}\verb@)@\\
\mbox{}\verb@  (trace@\\
\mbox{}\verb@    (recv @\hbox{$\langle\,$verification request\nobreak\ {\footnotesize \NWlink{nuweb6b}{6b}}$\,\rangle$}\verb@)@\\
\mbox{}\verb@    (recv @\hbox{$\langle\,$identity certificate\nobreak\ {\footnotesize \NWlink{nuweb8a}{8a}}$\,\rangle$}\verb@)@\\
\mbox{}\verb@    (send @\hbox{$\langle\,$verification query\nobreak\ {\footnotesize \NWlink{nuweb8b}{8b}}$\,\rangle$}\verb@)@\\
\mbox{}\verb@    (recv @\hbox{$\langle\,$report\nobreak\ {\footnotesize \NWlink{nuweb9b}{9b}}$\,\rangle$}\verb@)@\\
\mbox{}\verb@    (send @\hbox{$\langle\,$verification acceptance\nobreak\ {\footnotesize \NWlink{nuweb10b}{10b}}$\,\rangle$}\verb@))@\\
\mbox{}\verb@  (non-orig (invk hash) (privk e) (5 (ltk a a)))@\\
\mbox{}\verb@  (uniq-orig nv)@\\
\mbox{}\verb@  (annotations v@\\
\mbox{}\verb@    (1 (says e (id a i)))@\\
\mbox{}\verb@    (2 (ask r a j m))@\\
\mbox{}\verb@    (3 (says a (meas i nv j jo m p)))@\\
\mbox{}\verb@    (4 (approved r a nv))))@{\NWsep}
\end{list}
\vspace{-1ex}
\footnotesize\addtolength{\baselineskip}{-1ex}
\begin{list}{}{\setlength{\itemsep}{-\parsep}\setlength{\itemindent}{-\leftmargin}}
\item \NWtxtMacroRefIn\ \NWlink{nuweb23}{23}.
\end{list}
\end{minipage}\\[4ex]
\end{flushleft}
\begin{flushleft} \small
\begin{minipage}{\linewidth} \label{scrap20}
$\langle\,$verifier declarations\nobreak\ {\footnotesize \NWtarget{nuweb21b}{21b}}$\,\rangle\equiv$
\vspace{-1ex}
\begin{list}{}{} \item
\mbox{}\verb@(a v e s name) (r m p j jo text)@\\
\mbox{}\verb@(ns nv data) (hash i akey) (kp skey)@{\NWsep}
\end{list}
\vspace{-1ex}
\footnotesize\addtolength{\baselineskip}{-1ex}
\begin{list}{}{\setlength{\itemsep}{-\parsep}\setlength{\itemindent}{-\leftmargin}}
\item \NWtxtMacroRefIn\ \NWlink{nuweb21a}{21a}.
\end{list}
\end{minipage}\\[4ex]
\end{flushleft}
The channel key ${\ltk(A,A)}$ may be assumed to be non-originating
or not, depending on whether we wish to assume that both of the
VMs that communicate over this channel are uncompromised. If
${\ltk(A,A)}$ is non-originating, then two conclusions follow.  First,
the contents of communications along this channel are not disclosed.
Second, the endpoints of the channel, $C$ and $A$, are using the
channel only in accordance with this protocol.  Thus, any messages
sent or received over this channel are part of a regular,
protocol-respecting, execution of one of the roles.

If ${\ltk(A,A)}$ is not assumed to be non-originating, that means
that information on this
channel is available to an adversary, which could happen if one or
both of the endpoints are compromised.

An instance of a verifier role of height five has made a trust
decision---that the supplied measurement is acceptable.  The decision
is reflected by the non-origination assumption \texttt{(5 (ltk a a))}
in the role asserting that the channel key $\ltk(A,A)$ is uncompromised.

\begin{role}[Attester $A$]
\begin{eqnarray}
&\inbnd\ctoaa\\
&\outbnd B&\verifier(V)\land{}\label{eq:attester 1}\\
&&\meas(I,N_V,J,J_O,M,P)\nonumber
\end{eqnarray}
where $B=\blob$.  The initial theory of the attester defines the
$\verifier$ relation.
\end{role}

\begin{flushleft} \small
\begin{minipage}{\linewidth} \label{scrap21}
$\langle\,$attester role\nobreak\ {\footnotesize \NWtarget{nuweb22a}{22a}}$\,\rangle\equiv$
\vspace{-1ex}
\begin{list}{}{} \item
\mbox{}\verb@(defrole attester (vars @\hbox{$\langle\,$attester declarations\nobreak\ {\footnotesize \NWlink{nuweb22b}{22b}}$\,\rangle$}\verb@)@\\
\mbox{}\verb@  (trace@\\
\mbox{}\verb@    (recv @\hbox{$\langle\,$client query\nobreak\ {\footnotesize \NWlink{nuweb8d}{8d}}$\,\rangle$}\verb@)@\\
\mbox{}\verb@    (send @\hbox{$\langle\,$encoded report\nobreak\ {\footnotesize \NWlink{nuweb9a}{9a}}$\,\rangle$}\verb@))@\\
\mbox{}\verb@  (uniq-orig kp)@\\
\mbox{}\verb@  (non-orig (invk hash))@\\
\mbox{}\verb@  (annotations a@\\
\mbox{}\verb@    (1 (and (verifier v) (meas i nv j jo m p)))))@{\NWsep}
\end{list}
\vspace{-1ex}
\footnotesize\addtolength{\baselineskip}{-1ex}
\begin{list}{}{\setlength{\itemsep}{-\parsep}\setlength{\itemindent}{-\leftmargin}}
\item \NWtxtMacroRefIn\ \NWlink{nuweb23}{23}.
\end{list}
\end{minipage}\\[4ex]
\end{flushleft}
\begin{flushleft} \small
\begin{minipage}{\linewidth} \label{scrap22}
$\langle\,$attester declarations\nobreak\ {\footnotesize \NWtarget{nuweb22b}{22b}}$\,\rangle\equiv$
\vspace{-1ex}
\begin{list}{}{} \item
\mbox{}\verb@(a v s name) (r m p j jo text)@\\
\mbox{}\verb@(nv data) (hash i akey) (kp skey)@{\NWsep}
\end{list}
\vspace{-1ex}
\footnotesize\addtolength{\baselineskip}{-1ex}
\begin{list}{}{\setlength{\itemsep}{-\parsep}\setlength{\itemindent}{-\leftmargin}}
\item \NWtxtMacroRefIn\ \NWlink{nuweb22a}{22a}.
\end{list}
\end{minipage}\\[4ex]
\end{flushleft}
The CAVES protocol contains the five roles.
\begin{flushleft} \small
\begin{minipage}{\linewidth} \label{scrap23}
\verb@"caves.scm"@\nobreak\ {\footnotesize \NWtarget{nuweb23}{23} }$\equiv$
\vspace{-1ex}
\begin{list}{}{} \item
\mbox{}\verb@(defprotocol caves basic@\\
\mbox{}\verb@  @\hbox{$\langle\,$attester role\nobreak\ {\footnotesize \NWlink{nuweb22a}{22a}}$\,\rangle$}\verb@@\\
\mbox{}\verb@  @\hbox{$\langle\,$client role\nobreak\ {\footnotesize \NWlink{nuweb19a}{19a}}$\,\rangle$}\verb@@\\
\mbox{}\verb@  @\hbox{$\langle\,$server role\nobreak\ {\footnotesize \NWlink{nuweb20a}{20a}}$\,\rangle$}\verb@@\\
\mbox{}\verb@  @\hbox{$\langle\,$verifier role\nobreak\ {\footnotesize \NWlink{nuweb21a}{21a}}$\,\rangle$}\verb@@\\
\mbox{}\verb@  @\hbox{$\langle\,$enterprise privacy certificate authority role\nobreak\ {\footnotesize \NWlink{nuweb18}{18}}$\,\rangle$}\verb@)@{\NWsep}
\end{list}
\vspace{-1ex}
\footnotesize\addtolength{\baselineskip}{-1ex}
\begin{list}{}{\setlength{\itemsep}{-\parsep}\setlength{\itemindent}{-\leftmargin}}
\item \NWtxtFileDefBy\ \NWlink{nuweb17}{17}\NWlink{nuweb23}{, 23}\NWlink{nuweb24}{, 24}\NWlink{nuweb25a}{, 25a}\NWlink{nuweb25b}{b}\NWlink{nuweb25c}{c}\NWlink{nuweb26a}{, 26a}\NWlink{nuweb26b}{b}\NWlink{nuweb26c}{c}\NWlink{nuweb27a}{, 27a}\NWlink{nuweb27b}{b}.
\end{list}
\end{minipage}\\[4ex]
\end{flushleft}
\section{Scenarios}\label{sec:scenarios}

Most of the scenarios investigate the properties of the protocol
as seen from the point of view of one of the participants. The
scenario usually contains a regular strand for just one of the roles.
We can then ask an authentication question: if this strand
completes normally, are the other roles identified in this strand
actually present in the execution, and, if so, do they agree on
significant data such as keys and the identity of other parties?

We can also ask a confidentiality question for data that is uniquely
originated in the scenario role. A confidentiality question is tested
by adding a listener strand for the data in question.
Most of the attestation protocol principles are addressed by
authentication and confidentiality questions, and others are addressed
by annotations.

The ``bottom line'' of the protocol is the delivery of the data
$d$ to the client. The confidentiality of $d$ is an essential goal,
since the point of attestation is to ensure that the data goes only
to a client with acceptable measurements.

Origination assumptions play an important part in setting up and
interpreting scenarios.  For the CAVES protocol we pay special
attention to the origination assumption regarding the channel key
$\ltk(A,A)$.  The channel key ${\ltk(A,A)}$ may be assumed to be
non-originating or not, depending on whether we wish to assume that
both of the VMs (the client and the attester) that communicate over
this channel are uncompromised. If ${\ltk(A,A)}$ is non-originating,
then two conclusions follow.  First, the contents of communications
along this channel are not disclosed.  Second, the endpoints of the
channel, $C$ and $A$, which use this channel, act only in accordance
with this protocol.  Thus, any messages sent or received over this
channel are part of a regular, protocol-respecting, execution of one
of the roles.

If ${\ltk(A,A)}$ is not assumed to be non-originating, that means that
information on this channel may be available to an adversary.  If one
or both of the endpoints are compromised, or if the hypervisor exposes
the channel contents, this would occur.

The first scenario analyzes the protocol under the assumption of a
complete run of the verifier.  Since the verifier approved the release of
the data, the measurements reported by the attester are acceptable,
and the verifier can justify the assumption that the attester and
its associated client are regular. This is
indicated by assuming that \texttt{(ltk a a)} is non-originating by
assuming the verifier has height five.
In the CPSA results, we wish to confirm an authentication goal,
namely, that there was a regular attester strand agreeing on
the name \texttt{a} of the attester principal, and agreeing on the
measurements \texttt{p} and \texttt{jo} used to verify acceptability.

\begin{scenario}[Full Length Verifier]\label{scene:full length verifier}
\begin{flushleft} \small
\begin{minipage}{\linewidth} \label{scrap24}
\verb@"caves.scm"@\nobreak\ {\footnotesize \NWtarget{nuweb24}{24} }$\equiv$
\vspace{-1ex}
\begin{list}{}{} \item
\mbox{}\verb@(defskeleton caves@\\
\mbox{}\verb@  (vars (a v e s name))@\\
\mbox{}\verb@  (defstrand verifier 5 (a a) (v v) (e e) (s s))@\\
\mbox{}\verb@  (non-orig (privk v) (privk e) (privk s)))@{\NWsep}
\end{list}
\vspace{-1ex}
\footnotesize\addtolength{\baselineskip}{-1ex}
\begin{list}{}{\setlength{\itemsep}{-\parsep}\setlength{\itemindent}{-\leftmargin}}
\item \NWtxtFileDefBy\ \NWlink{nuweb17}{17}\NWlink{nuweb23}{, 23}\NWlink{nuweb24}{, 24}\NWlink{nuweb25a}{, 25a}\NWlink{nuweb25b}{b}\NWlink{nuweb25c}{c}\NWlink{nuweb26a}{, 26a}\NWlink{nuweb26b}{b}\NWlink{nuweb26c}{c}\NWlink{nuweb27a}{, 27a}\NWlink{nuweb27b}{b}.
\end{list}
\end{minipage}\\[4ex]
\end{flushleft}
\end{scenario}

CPSA finds a shape with the normal five strands, where the height of
each strand is the full length with the exception of the client
(height 5) and the server (height 4), since the verifier has no
information about what happened after its own last message.  The
values on which the strands agree is as expected.

The fresh information principle is respected in this scenario because
the verifier generated the nonce \texttt{nv} before the measurement
blob was created.

To illustrate the importance of the non-origination assumption on
\texttt{(ltk a a)} that was added because the verifier had height
five, a similar scenario is run where the verifier has height four.

\begin{scenario}[Verifier Before Decision]
\label{scene:verifier before decision}
\begin{flushleft} \small
\begin{minipage}{\linewidth} \label{scrap25}
\verb@"caves.scm"@\nobreak\ {\footnotesize \NWtarget{nuweb25a}{25a} }$\equiv$
\vspace{-1ex}
\begin{list}{}{} \item
\mbox{}\verb@(defskeleton caves@\\
\mbox{}\verb@  (vars (e s name))@\\
\mbox{}\verb@  (defstrand verifier 4 (e e) (s s))@\\
\mbox{}\verb@  (non-orig (privk e) (privk s)))@{\NWsep}
\end{list}
\vspace{-1ex}
\footnotesize\addtolength{\baselineskip}{-1ex}
\begin{list}{}{\setlength{\itemsep}{-\parsep}\setlength{\itemindent}{-\leftmargin}}
\item \NWtxtFileDefBy\ \NWlink{nuweb17}{17}\NWlink{nuweb23}{, 23}\NWlink{nuweb24}{, 24}\NWlink{nuweb25a}{, 25a}\NWlink{nuweb25b}{b}\NWlink{nuweb25c}{c}\NWlink{nuweb26a}{, 26a}\NWlink{nuweb26b}{b}\NWlink{nuweb26c}{c}\NWlink{nuweb27a}{, 27a}\NWlink{nuweb27b}{b}.
\end{list}
\end{minipage}\\[4ex]
\end{flushleft}
\end{scenario}

CPSA finds a shape with four strands---it cannot infer the existence
of a client strand. Thus, the session key \texttt{k} received
by the server may have been generated by the adversary.

The following scenarios are from the point of view of the
attester. First, suppose that \texttt{(ltk a a)} is
non-originating.

\begin{scenario}[Full Length Attester]
\begin{flushleft} \small
\begin{minipage}{\linewidth} \label{scrap26}
\verb@"caves.scm"@\nobreak\ {\footnotesize \NWtarget{nuweb25b}{25b} }$\equiv$
\vspace{-1ex}
\begin{list}{}{} \item
\mbox{}\verb@(defskeleton caves@\\
\mbox{}\verb@  (vars (a v name))@\\
\mbox{}\verb@  (defstrand attester 2 (a a) (v v))@\\
\mbox{}\verb@  (non-orig (privk v) (ltk a a)))@{\NWsep}
\end{list}
\vspace{-1ex}
\footnotesize\addtolength{\baselineskip}{-1ex}
\begin{list}{}{\setlength{\itemsep}{-\parsep}\setlength{\itemindent}{-\leftmargin}}
\item \NWtxtFileDefBy\ \NWlink{nuweb17}{17}\NWlink{nuweb23}{, 23}\NWlink{nuweb24}{, 24}\NWlink{nuweb25a}{, 25a}\NWlink{nuweb25b}{b}\NWlink{nuweb25c}{c}\NWlink{nuweb26a}{, 26a}\NWlink{nuweb26b}{b}\NWlink{nuweb26c}{c}\NWlink{nuweb27a}{, 27a}\NWlink{nuweb27b}{b}.
\end{list}
\end{minipage}\\[4ex]
\end{flushleft}
\end{scenario}

CPSA finds a shape with the attester and a regular client, but nothing else.
Now, what if \texttt{(ltk a a)} is not non-originating?

\begin{scenario}[Full Length Attester, Compromised Client]
\begin{flushleft} \small
\begin{minipage}{\linewidth} \label{scrap27}
\verb@"caves.scm"@\nobreak\ {\footnotesize \NWtarget{nuweb25c}{25c} }$\equiv$
\vspace{-1ex}
\begin{list}{}{} \item
\mbox{}\verb@(defskeleton caves@\\
\mbox{}\verb@  (vars (a v name))@\\
\mbox{}\verb@  (defstrand attester 2 (a a) (v v))@\\
\mbox{}\verb@  (non-orig (privk v)))@{\NWsep}
\end{list}
\vspace{-1ex}
\footnotesize\addtolength{\baselineskip}{-1ex}
\begin{list}{}{\setlength{\itemsep}{-\parsep}\setlength{\itemindent}{-\leftmargin}}
\item \NWtxtFileDefBy\ \NWlink{nuweb17}{17}\NWlink{nuweb23}{, 23}\NWlink{nuweb24}{, 24}\NWlink{nuweb25a}{, 25a}\NWlink{nuweb25b}{b}\NWlink{nuweb25c}{c}\NWlink{nuweb26a}{, 26a}\NWlink{nuweb26b}{b}\NWlink{nuweb26c}{c}\NWlink{nuweb27a}{, 27a}\NWlink{nuweb27b}{b}.
\end{list}
\end{minipage}\\[4ex]
\end{flushleft}
\end{scenario}

In this case, CPSA finds a shape with the attester only.
There is still a client being measured, but the client
may not be regular.

One of the attestation principles is constrained disclosure.
We can test this by adding a listener for the measurements
\texttt{jo} or \texttt{p}. The following is
the test for \texttt{jo}. Note that \texttt{jo} must be
assumed uniquely originating in order for the confidentiality
test with the listener to be meaningful.

\begin{scenario}[Full Length Attester, Compromised Client,
Listener for \texttt{jo} ]\label{scene:full jo listener}
\begin{flushleft} \small
\begin{minipage}{\linewidth} \label{scrap28}
\verb@"caves.scm"@\nobreak\ {\footnotesize \NWtarget{nuweb26a}{26a} }$\equiv$
\vspace{-1ex}
\begin{list}{}{} \item
\mbox{}\verb@(defskeleton caves@\\
\mbox{}\verb@  (vars (a v name) (jo text))@\\
\mbox{}\verb@  (defstrand attester 2 (a a) (v v) (jo jo))@\\
\mbox{}\verb@  (deflistener jo)@\\
\mbox{}\verb@  (uniq-orig jo)@\\
\mbox{}\verb@  (non-orig (privk v)))@{\NWsep}
\end{list}
\vspace{-1ex}
\footnotesize\addtolength{\baselineskip}{-1ex}
\begin{list}{}{\setlength{\itemsep}{-\parsep}\setlength{\itemindent}{-\leftmargin}}
\item \NWtxtFileDefBy\ \NWlink{nuweb17}{17}\NWlink{nuweb23}{, 23}\NWlink{nuweb24}{, 24}\NWlink{nuweb25a}{, 25a}\NWlink{nuweb25b}{b}\NWlink{nuweb25c}{c}\NWlink{nuweb26a}{, 26a}\NWlink{nuweb26b}{b}\NWlink{nuweb26c}{c}\NWlink{nuweb27a}{, 27a}\NWlink{nuweb27b}{b}.
\end{list}
\end{minipage}\\[4ex]
\end{flushleft}
\end{scenario}

CPSA does not find a realized shape with the listener,
confirming that the measurement is kept confidential.
A similar test for \texttt{p} shows that the PCR vector
is kept confidential as well.

\begin{scenario}[Full Length Attester,
Compromised Client, Listener for \texttt{p}]
\begin{flushleft} \small
\begin{minipage}{\linewidth} \label{scrap29}
\verb@"caves.scm"@\nobreak\ {\footnotesize \NWtarget{nuweb26b}{26b} }$\equiv$
\vspace{-1ex}
\begin{list}{}{} \item
\mbox{}\verb@(defskeleton caves@\\
\mbox{}\verb@  (vars (a v name) (p text))@\\
\mbox{}\verb@  (defstrand attester 2 (a a) (v v) (p p))@\\
\mbox{}\verb@  (deflistener p)@\\
\mbox{}\verb@  (uniq-orig p)@\\
\mbox{}\verb@  (non-orig (privk v)))@{\NWsep}
\end{list}
\vspace{-1ex}
\footnotesize\addtolength{\baselineskip}{-1ex}
\begin{list}{}{\setlength{\itemsep}{-\parsep}\setlength{\itemindent}{-\leftmargin}}
\item \NWtxtFileDefBy\ \NWlink{nuweb17}{17}\NWlink{nuweb23}{, 23}\NWlink{nuweb24}{, 24}\NWlink{nuweb25a}{, 25a}\NWlink{nuweb25b}{b}\NWlink{nuweb25c}{c}\NWlink{nuweb26a}{, 26a}\NWlink{nuweb26b}{b}\NWlink{nuweb26c}{c}\NWlink{nuweb27a}{, 27a}\NWlink{nuweb27b}{b}.
\end{list}
\end{minipage}\\[4ex]
\end{flushleft}
\end{scenario}

The next scenario, with a full-length server, should yield a
normal execution of the protocol.

\begin{scenario}[Full Length Server]\label{scene:full length server}
\begin{flushleft} \small
\begin{minipage}{\linewidth} \label{scrap30}
\verb@"caves.scm"@\nobreak\ {\footnotesize \NWtarget{nuweb26c}{26c} }$\equiv$
\vspace{-1ex}
\begin{list}{}{} \item
\mbox{}\verb@(defskeleton caves@\\
\mbox{}\verb@  (vars (v s name))@\\
\mbox{}\verb@  (defstrand server 8 (s s) (v v))@\\
\mbox{}\verb@  (non-orig (privk s) (privk v)))@{\NWsep}
\end{list}
\vspace{-1ex}
\footnotesize\addtolength{\baselineskip}{-1ex}
\begin{list}{}{\setlength{\itemsep}{-\parsep}\setlength{\itemindent}{-\leftmargin}}
\item \NWtxtFileDefBy\ \NWlink{nuweb17}{17}\NWlink{nuweb23}{, 23}\NWlink{nuweb24}{, 24}\NWlink{nuweb25a}{, 25a}\NWlink{nuweb25b}{b}\NWlink{nuweb25c}{c}\NWlink{nuweb26a}{, 26a}\NWlink{nuweb26b}{b}\NWlink{nuweb26c}{c}\NWlink{nuweb27a}{, 27a}\NWlink{nuweb27b}{b}.
\end{list}
\end{minipage}\\[4ex]
\end{flushleft}
\end{scenario}

CPSA finds the expected execution.

To this scenario, we add a listener for \texttt{d} to
check that the data is not compromised.

\begin{scenario}[$D$ Kept Secret in Full Length Server]
\label{scene:d kept secret in server}
\begin{flushleft} \small
\begin{minipage}{\linewidth} \label{scrap31}
\verb@"caves.scm"@\nobreak\ {\footnotesize \NWtarget{nuweb27a}{27a} }$\equiv$
\vspace{-1ex}
\begin{list}{}{} \item
\mbox{}\verb@(defskeleton caves@\\
\mbox{}\verb@  (vars (v s name) (d text))@\\
\mbox{}\verb@  (defstrand server 8 (s s) (v v) (d d))@\\
\mbox{}\verb@  (deflistener d)@\\
\mbox{}\verb@  (uniq-orig d)@\\
\mbox{}\verb@  (non-orig (privk s) (privk v)))@{\NWsep}
\end{list}
\vspace{-1ex}
\footnotesize\addtolength{\baselineskip}{-1ex}
\begin{list}{}{\setlength{\itemsep}{-\parsep}\setlength{\itemindent}{-\leftmargin}}
\item \NWtxtFileDefBy\ \NWlink{nuweb17}{17}\NWlink{nuweb23}{, 23}\NWlink{nuweb24}{, 24}\NWlink{nuweb25a}{, 25a}\NWlink{nuweb25b}{b}\NWlink{nuweb25c}{c}\NWlink{nuweb26a}{, 26a}\NWlink{nuweb26b}{b}\NWlink{nuweb26c}{c}\NWlink{nuweb27a}{, 27a}\NWlink{nuweb27b}{b}.
\end{list}
\end{minipage}\\[4ex]
\end{flushleft}
\end{scenario}

CPSA finds no shape, so the secret is kept.

The client has an authentication goal to ensure that the
data it receives comes from the chosen server.
Since the client is regular, we assume a safe channel key,
and since the client selects the server, we assume the
server's public key is safe as well.

\begin{scenario}[Full Length Client]\label{scene:full length client}
\begin{flushleft} \small
\begin{minipage}{\linewidth} \label{scrap32}
\verb@"caves.scm"@\nobreak\ {\footnotesize \NWtarget{nuweb27b}{27b} }$\equiv$
\vspace{-1ex}
\begin{list}{}{} \item
\mbox{}\verb@(defskeleton caves@\\
\mbox{}\verb@  (vars (a v s name) (k skey))@\\
\mbox{}\verb@  (defstrand client 6 (v v) (a a) (s s) (k k))@\\
\mbox{}\verb@  (non-orig (privk s) (privk v) (ltk a a)))@{\NWsep}
\end{list}
\vspace{-1ex}
\footnotesize\addtolength{\baselineskip}{-1ex}
\begin{list}{}{\setlength{\itemsep}{-\parsep}\setlength{\itemindent}{-\leftmargin}}
\item \NWtxtFileDefBy\ \NWlink{nuweb17}{17}\NWlink{nuweb23}{, 23}\NWlink{nuweb24}{, 24}\NWlink{nuweb25a}{, 25a}\NWlink{nuweb25b}{b}\NWlink{nuweb25c}{c}\NWlink{nuweb26a}{, 26a}\NWlink{nuweb26b}{b}\NWlink{nuweb26c}{c}\NWlink{nuweb27a}{, 27a}\NWlink{nuweb27b}{b}.
\end{list}
\end{minipage}\\[4ex]
\end{flushleft}
\end{scenario}

CPSA produces a normal shape as in Scenario~\ref{scene:full length server}
In particular, the client and server agree on their identities and on
the data sent to the client.

\section{Trust Argument}\label{sec:trust argument}

Figure~\ref{fig:rg} shows the rely and guarantee formulas together as
specified in Section~\ref{sec:roles}. These formulas also appear in
the CPSA output for Scenario~\ref{scene:full length server}, where the
formulas are instantiated in a way that is consistent across strands.

The purpose of rely and guarantee formulas is to supply
application-specific context for the data sent in messages.  When the
server chooses a verifier to contact, for example, the guarantee of
$\verifier(V)$ is what identifies $V$ as the name of a principal
authorized to play a verifier role. The implementor or administrator
of the protocol has the responsibility to ensure in some way, not
specified in the protocol itself, that only appropriate verifier
principals are chosen. Similar remarks apply to other guarantees, such
as those that choose or evaluate measurements for acceptability.

For each non-trivial formula that annotates the shape,
Table~\ref{fig:rg} contains a row.  The row entries are (1) the node
of the formula, (2) the principal associated with its strand, and (3)
the formula.  A node with a plus sign is a transmission node and its
formula is guaranteed before the message is sent.  A node with a minus
sign is a reception node, and when the shape is sound, its formula can
be relied upon as a result of receiving the message.  CPSA does not
itself check soundness, but it does generate proof obligations stating
that a rely formula $\psi$ follows from the formulas $A\says\phi$,
where each formula $\phi$ was guaranteed by principal $A$ at some
transmission node that precedes $\psi$ in the node ordering.  An
external theorem prover could be used to check that each such formula
$(\bigwedge A\says\phi)\supset \psi$ is logically valid.

\begin{figure}
$$\xymatrix@R=1.8ex{
\txt{\strut Attester}&\txt{\strut Client}&\txt{\strut Server}&\txt{\strut Verifier}&\txt{\strut EPCA}\\
&1,0\ar@{=>}[dddd]\ar[r]&2,0\ar@{=>}[d]&&\\
&&2,1\ar@{=>}[dd]\ar[r]&3,0\ar@{=>}[d]&\\
&&&3,1\ar@{=>}[d]&4,0\ar[l]\\
&&2,2\ar@{=>}[d]&3,2\ar@{=>}[ddddd]\ar[l]&\\
&1,1\ar@{=>}[d]&2,3\ar@{=>}[ddd]\ar[l]&&\\
0,0\ar@{=>}[d]&1,2\ar@{=>}[d]\ar[l]&&&\\
0,1\ar[r]&1,3\ar@{=>}[d]&&&\\
&1,4\ar@{=>}[ddd]\ar[r]&2,4\ar@{=>}[d]&&\\
&&2,5\ar@{=>}[d]\ar[r]&3,3\ar@{=>}[d]&\\
&&2,6\ar@{=>}[d]&3,4\ar[l]&\\
&1,5&2,7\ar[l]&&}$$
\begin{eqnarray*}
\outbnd(2,1)&S&\verifier(V)\\
\outbnd(4,0)&E&\id(A,I)\\
\inbnd(3,1)&V&E\says\id(A,I)\\
\outbnd(3,2)&V&\ask(R,A,J,M)\\
\outbnd(0,1)&A&\meas(I,N_V,J,J_O,M,P)\\
\inbnd(3,3)&V&A\says\meas(I,N_V,J,J_O,M,P)\\
\outbnd(3,4)&V&\approved(R,A,N_V)\\
\inbnd(2,6)&S&V\says\approved(R,A,N_V)\\
\outbnd(2,7)&S&\resource(R,D) \wedge \approved(R,A,N_V)\\
\inbnd(1,5)&C&S\says\resource(R,D)
\end{eqnarray*}
\caption{CAVE Rely-Guarantee Formulas}\label{fig:rg}
\end{figure}

In all shapes found for CAVES, the soundness proof obligations are
trivially true.  For example, the guarantee made by the attester at
node~$(0,1)$ justifies the reliance by the verifier on the formula at
node~$(3,3)$.

From the client's point of view, the client wants to know that the
received data was sent by the specified server for its resource
request.  Scenario~\ref{scene:full length client} shows that a regular
server exists and agrees with the client on their identities and the
parameter $D$, as well as the request $R$, but the client depends on
the rely formula at node~$(1,5)$ in Figure~\ref{fig:rg} to decide that
$D$ is the requested data.

A goal of the protocol from the server's point of view is that only a
valid client receives the resource data.  There are three questions
buried in this concern. One is that the client is ``valid''. Another
is that only the identified client receives the data. And, finally,
that content $D$ of the message is in fact the requested resource
data.

We have already tested, in Scenario~\ref{scene:d kept secret in
  server}, that $D$ is not compromised by the adversary, so it is
received only by a regular participant.  But is that participant
``valid''? And, if it is, we must still ask why we think that $D$ is
the resource data. The answer to the latter question is the guarantee
$\resource(R,D)$ in node~$+(2,7).$ Of course, only the implementor
knows what kind of database operation or computation was called for to
produce that data.

Why is the recipient of the data ``valid''? In this case, ``valid'' is
defined by the guarantee $\approved(R,A,N_V)$ at node~$+(2,7)$.  This
guarantee is justified by Inference Rule~\ref{eq:trust verifier}
on Page~\pageref{eq:trust verifier} which depends on hypotheses
provided by annotations at nodes~$+(2,1)$ and $-(2,6)$.

The validity of guarantees justified by inference rules depends on the
validity of the inference rules and the chain of deductions using
them.  This process is not checked by CPSA, but the necessary formulas
and rules are made visible, and other tools can be brought to bear to
check them.

The rest of the argument is summarized here to suggest why several of
the other rules and guarantees are needed. Most of the logical
activity occurs in the verifier.

The verifier guarantees the client associated with verification
session~$N_V$ is valid before sending the message at node~$(3,4)$
using Inference Rule~\ref{eq:approved rule} on
Page~\pageref{eq:approved rule}.  To use this rule, the verifier must
obtain the attestation identity key of the attester, receive a
measurement from the attester, and check that the measurement values
are correct by consulting the $\ok$ relation in its initial theory.

At node~$(3,0)$, the verifier receives the distinguished name~$A$ used
in the attestation identity certificate containing the attester's
identity key.  After receiving the certificate at node~$(3,1)$, the
verifier relies on the fact that the enterprise privacy certificate
authority~$E$ says that~$I$ is~$A$'s identity key.  After consulting
the $\epca$ relation in its initial theory, and using the Inference
Rule~\ref{eq:trust epca}, the verifier guarantees that~$I$ is~$A$'s
identity key at node~$(3,4)$.

The verifier relies on the fact that attester~$A$ claims it provides
some measurements of the client after receiving the message at
node~$(3,3)$.  For this shape, the measured values agree with the
expected values, otherwise, the message transmission at node~$(3,4)$
would have been forbidden because its formula could not be guaranteed.

The guarantee at node~$(3,2)$ and Inference Rule~\ref{eq:ask}
select a PCR mask~$M$ and a measurement query~$J$ that allows the
check of the measurements before node~$(3,4)$ using Inference
Rule~\ref{eq:approved rule}.

The last non-trivial formula to be discussed is the guarantee made by
the attester.  It asserts that~$V$ is a verifier, and for verification
session~$N_V$, the requested measurement result is~$J_O$ for
query~$J$, and the requested PCR vector is~$P$ for PCR mask~$M$.  It
must guarantee that~$V$ is verifier so as to ensure the measurements
are available only to a party authorized to view them.

The server delegates to the verifier the decision to release data to
the client.  The verifier approves the release by sending its last
message only if the attester provides credible evidence that the
client is valid.  The rely-guarantee formulas formalize this trust
argument.

\section{High-Level Attestation Goals}

\begin{description}
  \item[Fresh information] Evidence is up to date because the
  attester's nodes are between the verifier nodes~$(3,2)$ and
  $(3,3)$.  (See e.g.~Scenario~\ref{scene:full length verifier}.)
  \item[Comprehensive information] The query $J$ may be chosen to
  provide sufficient insight into the state of the client $C$, as
  reported in output $J_O$, to make an adjudication that $C$ will
  behave regularly.
  \item[Constrained disclosure] The measurement is not available to
  the client and the server, and the attester authenticates the
  verifier before sending out the measurement.  (See
  e.g.~Scenario~\ref{scene:full jo listener}.)
\item[Semantic explicitness] The use of the rely-guarantee method
  provides formal description of the access decision.
\item[Trustworthy mechanism] The TPM provides a root of trust for
  reporting via the TPM quote included in the message sent at
  node~$(0,1)$.
\end{description}

\section{Protocol Development History}

An earlier version of this protocol omitted~$S$ and~$N_V$ from
Message~\ref{eq:attester request} in Figure~\ref{fig:caves} and the
contents of the blob~$B$.  A CPSA analysis of this version revealed a
man-in-the-middle attack due to a possible disagreement on the
identity of the client between the server and the attester. Some
protocol modifications were tried, and several CPSA runs were
made. For a sample of the analysis, consider the following result of a
run in which the only difference from the current protocol is the
absence of $S$ from the attester messages. The scenario had a
full-length server and a listener for $D$, and the channel key was
assumed non-originating.  This run yielded a shape with all roles
represented, and with the following strand mappings:
\begin{verbatim}
  (defstrand server 8 (b b) (r r) (m m) (j j) (d d) (ns ns)
    (nv nv) (a a) (v v) (s s) (k k))
  (deflistener d)
  (defstrand epca 1 (a a) (e e) (i i))
  (defstrand verifier 5 (r r) (m m) (p p) (j j) (jo jo) (ns ns)
    (nv nv) (a a) (v v) (e e) (s s) (hash hash) (i i))
  (defstrand attester 2 (r r) (m m) (p p) (j j) (jo jo) (nv nv)
    (a a) (v v) (hash hash) (i i))
  (defstrand client 3 (r r) (m m) (j j) (nv nv) (s s-0) (c c)
    (a a) (v v) (k k-0))
\end{verbatim}
The anomaly here is that the client disagrees with the server on the
server identity \texttt{s-0} and the session key \texttt{k-0}.  The
data \texttt{d} was compromised because the server received a
different session key \texttt{k} from the adversary. For the attack to
work, the adversary must possess the private key of \texttt{s-0}. That
is, the client is communicating with a malicious server.

Omitting the occurrence of~$N_V$ outside the blob in
Message~\ref{eq:attester response} in Figure~\ref{fig:caves} allows
shapes with two clients.  Inserting~$N_V$ outside the blob was the
final protocol design step.

\section{Conclusion}

In this report, we have highlighted the role that CPSA analysis has
played in refining and justifying the design of the CAVES
protocol. Shapes analysis without annotations allowed us to prove
authentication and confidentiality properties.  The rely and guarantee
annotations bridge the gap between message behavior and its
application-specific semantics. CPSA also allowed us to check the
soundness of the reliance by one participant on another participant's
guarantees.

The process of representing the protocol in the CPSA language for
purposes of analysis led to formulation of generally applicable
techniques for modeling such phenomena as private channels---using an
auxiliary long-term key like $\ltk(A,A)$---and hash functions.

We hope that the use of Nuweb to weave together the specification text
and the associated exposition has demonstrated how we can effectively
combine versatility of narrative style with a tight coupling to the
actual executable text.

This analysis of CAVES is one step in a larger plan to analyze related
protocols such as the remeasurement protocol and a more general
protocol that can negotiate and launch a selection of measurement and
attestation functions.  We also plan to address other questions
associated with logical annotations, such as how inference rules may
be justified by analysis based on models of the TRP itself and its
components.

\bibliography{caves}
\bibliographystyle{plain}

\ifshowresults
\appendix

\section{Results}

What follows is the verbatim output of CPSA except that duplicate
protocols listings have been removed and scenarios section titles
added.  If available, the XHTML rendering of this information should
be preferred as hyperlinks and SVG diagrams enhance the presentation.


\begingroup\small
\begin{verbatim}
(comment "CPSA 2.2.0")
(comment "Annotated skeletons")
(comment "CPSA 2.2.0")
(comment "Extracted shapes")
(herald "CAVES Attestation Protocol" (bound 12) (check-nonces))
(comment "CPSA 2.2.0")
(comment "All input read")
(comment "Strand count bounded at 12")
(comment "Nonces checked first")

(defprotocol caves basic
  (defrole attester
    (vars (a v s name) (r m p j jo text) (nv data) (hash i akey)
      (kp skey))
    (trace (recv (enc s nv j v m r (ltk a a)))
      (send
        (enc kp nv
          (enc kp s jo m p
            (enc
              (enc "hash" (enc "hash" a v r nv j jo hash) m p
                hash) (invk i)) (pubk v)) (ltk a a))))
    (non-orig (invk hash))
    (uniq-orig kp)
    (annotations a (1 (and (verifier v) (meas i nv j jo m p)))))
  (defrole client
    (vars (c a v s name) (r m j d text) (nv data) (k kp skey)
      (b mesg))
    (trace (send (cat c (enc r a k (pubk s))))
      (recv (enc nv j v m k)) (send (enc s nv j v m r (ltk a a)))
      (recv (enc kp nv b (ltk a a))) (send b)
      (recv (enc "data" d kp)))
    (uniq-orig k)
    (annotations c (5 (says s (resource r d)))))
  (defrole server
    (vars (a v s name) (r m j d text) (ns nv data) (k kp skey)
      (b mesg))
    (trace (recv (enc r a k (pubk s)))
      (send (enc s r a ns (pubk v)))
      (recv (enc nv j v ns m (pubk s))) (send (enc nv j v m k))
      (recv b) (send b) (recv (enc "valid" kp ns (pubk s)))
      (send (enc "data" d kp)))
    (uniq-orig ns)
    (annotations s (1 (verifier v)) (6 (says v (approved r a nv)))
      (7 (and (approved r a nv) (resource r d)))))
  (defrole verifier
    (vars (a v e s name) (r m p j jo text) (ns nv data)
      (hash i akey) (kp skey))
    (trace (recv (enc s r a ns (pubk v)))
      (recv (enc "cert" a i e (privk e)))
      (send (enc nv j v ns m (pubk s)))
      (recv
        (enc kp s jo m p
          (enc
            (enc "hash" (enc "hash" a v r nv j jo hash) m p hash)
            (invk i)) (pubk v)))
      (send (enc "valid" kp ns (pubk s))))
    (non-orig (invk hash) (privk e) (5 (ltk a a)))
    (uniq-orig nv)
    (annotations v (1 (says e (id a i))) (2 (ask r a j m))
      (3 (says a (meas i nv j jo m p))) (4 (approved r a nv))))
  (defrole epca
    (vars (a e name) (i akey))
    (trace (send (enc "cert" a i e (privk e))))
    (non-orig (invk i))
    (annotations e (0 (id a i)))))
\end{verbatim}

\subsection{Scenario}

$$\xymatrix{
\txt{\strut verifier}\\
\bullet\ar@{=>}[d]\\
\bullet\ar@{=>}[d]\\
\bullet\ar@{=>}[d]\\
\bullet\ar@{=>}[d]\\
\bullet}$$

\begin{verbatim}
(defskeleton caves
  (vars (r m p j jo text) (ns nv data) (a v e s name) (kp skey)
    (hash i akey))
  (defstrand verifier 5 (r r) (m m) (p p) (j j) (jo jo) (ns ns)
    (nv nv) (a a) (v v) (e e) (s s) (kp kp) (hash hash) (i i))
  (non-orig (ltk a a) (invk hash) (privk v) (privk e) (privk s))
  (uniq-orig nv)
  (traces
    ((recv (enc s r a ns (pubk v)))
      (recv (enc "cert" a i e (privk e)))
      (send (enc nv j v ns m (pubk s)))
      (recv
        (enc kp s jo m p
          (enc
            (enc "hash" (enc "hash" a v r nv j jo hash) m p hash)
            (invk i)) (pubk v)))
      (send (enc "valid" kp ns (pubk s)))))
  (label 0)
  (unrealized (0 1) (0 3))
  (comment "1 in cohort - 1 not yet seen"))
\end{verbatim}

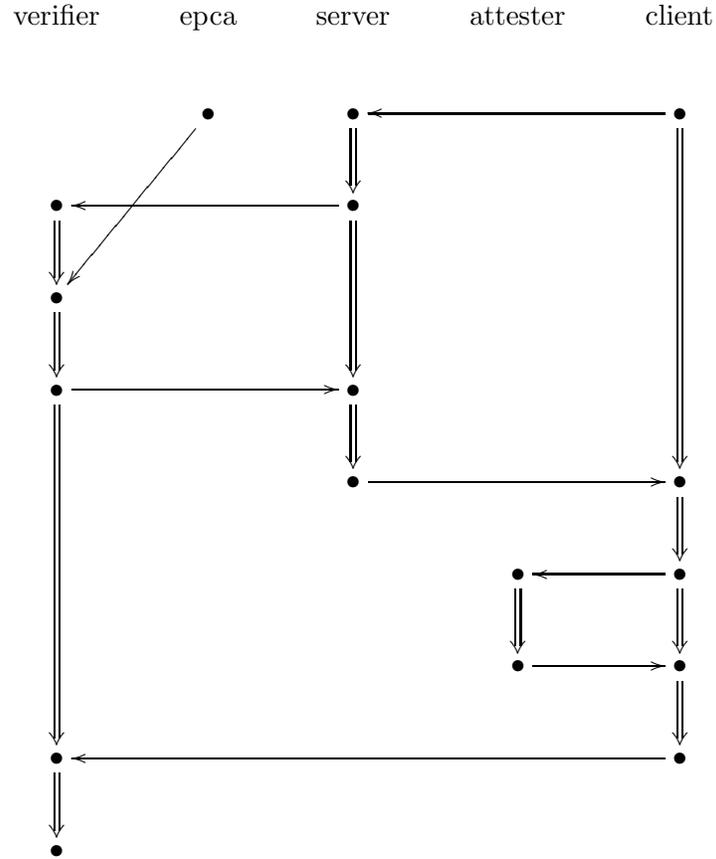
$$\xymatrix{
\txt{\strut verifier}&\txt{\strut epca}&\txt{\strut server}&\txt{\strut attester}&\txt{\strut client}\\
&\bullet\ar[ldd]&\bullet\ar@{=>}[d]&&\bullet\ar@{=>}[dddd]\ar[ll]\\
\bullet\ar@{=>}[d]&&\bullet\ar@{=>}[dd]\ar[ll]&&\\
\bullet\ar@{=>}[d]&&&&\\
\bullet\ar@{=>}[dddd]\ar[rr]&&\bullet\ar@{=>}[d]&&\\
&&\bullet\ar[rr]&&\bullet\ar@{=>}[d]\\
&&&\bullet\ar@{=>}[d]&\bullet\ar@{=>}[d]\ar[l]\\
&&&\bullet\ar[r]&\bullet\ar@{=>}[d]\\
\bullet\ar@{=>}[d]&&&&\bullet\ar[llll]\\
\bullet&&&&}$$

\begin{verbatim}
(defskeleton caves
  (vars (m p j jo r text) (ns nv data) (v e a c s name)
    (k kp skey) (hash i akey))
  (defstrand verifier 5 (r r) (m m) (p p) (j j) (jo jo) (ns ns)
    (nv nv) (a a) (v v) (e e) (s s) (kp kp) (hash hash) (i i))
  (defstrand epca 1 (a a) (e e) (i i))
  (defstrand server 4 (r r) (m m) (j j) (ns ns) (nv nv) (a a)
    (v v) (s s) (k k))
  (defstrand attester 2 (r r) (m m) (p p) (j j) (jo jo) (nv nv)
    (a a) (v v) (s s) (kp kp) (hash hash) (i i))
  (defstrand client 5
    (b
      (enc kp s jo m p
        (enc (enc "hash" (enc "hash" a v r nv j jo hash) m p hash)
          (invk i)) (pubk v))) (r r) (m m) (j j) (nv nv) (c c)
    (a a) (v v) (s s) (k k) (kp kp))
  (precedes ((0 2) (2 2)) ((1 0) (0 1)) ((2 1) (0 0))
    ((2 3) (4 1)) ((3 1) (4 3)) ((4 0) (2 0)) ((4 2) (3 0))
    ((4 4) (0 3)))
  (non-orig (ltk a a) (invk hash) (invk i) (privk v) (privk e)
    (privk s))
  (uniq-orig ns nv k kp)
  (operation nonce-test
    (contracted (r-0 r) (s-0 s) (c-0 c) (v-0 v) (m-0 m) (j-0 j)
      (k-0 k)) nv (5 1) (enc nv j v m k)
    (enc nv j v ns m (pubk s)))
  (traces
    ((recv (enc s r a ns (pubk v)))
      (recv (enc "cert" a i e (privk e)))
      (send (enc nv j v ns m (pubk s)))
      (recv
        (enc kp s jo m p
          (enc
            (enc "hash" (enc "hash" a v r nv j jo hash) m p hash)
            (invk i)) (pubk v)))
      (send (enc "valid" kp ns (pubk s))))
    ((send (enc "cert" a i e (privk e))))
    ((recv (enc r a k (pubk s))) (send (enc s r a ns (pubk v)))
      (recv (enc nv j v ns m (pubk s))) (send (enc nv j v m k)))
    ((recv (enc s nv j v m r (ltk a a)))
      (send
        (enc kp nv
          (enc kp s jo m p
            (enc
              (enc "hash" (enc "hash" a v r nv j jo hash) m p
                hash) (invk i)) (pubk v)) (ltk a a))))
    ((send (cat c (enc r a k (pubk s)))) (recv (enc nv j v m k))
      (send (enc s nv j v m r (ltk a a)))
      (recv
        (enc kp nv
          (enc kp s jo m p
            (enc
              (enc "hash" (enc "hash" a v r nv j jo hash) m p
                hash) (invk i)) (pubk v)) (ltk a a)))
      (send
        (enc kp s jo m p
          (enc
            (enc "hash" (enc "hash" a v r nv j jo hash) m p hash)
            (invk i)) (pubk v)))))
  (label 29)
  (parent 0)
  (unrealized)
  (shape)
  (annotations ((0 1) v (says e (id a i))) ((0 2) v (ask r a j m))
    ((0 3) v (says a (meas i nv j jo m p)))
    ((0 4) v (approved r a nv)) ((1 0) e (id a i))
    ((2 1) s (verifier v))
    ((3 1) a (and (verifier v) (meas i nv j jo m p))))
  (obligations
    ((0 1) v
      (implies (says e (id a i)) (says s (verifier v))
        (says e (id a i))))
    ((0 3) v
      (implies (ask r a j m) (says e (id a i))
        (says s (verifier v))
        (says a (and (verifier v) (meas i nv j jo m p)))
        (says a (meas i nv j jo m p))))))
\end{verbatim}

\subsection{Scenario}

$$\xymatrix{
\txt{\strut verifier}\\
\bullet\ar@{=>}[d]\\
\bullet\ar@{=>}[d]\\
\bullet\ar@{=>}[d]\\
\bullet}$$

\begin{verbatim}
(defskeleton caves
  (vars (r m p j jo text) (ns nv data) (e s a v name) (kp skey)
    (hash i akey))
  (defstrand verifier 4 (r r) (m m) (p p) (j j) (jo jo) (ns ns)
    (nv nv) (a a) (v v) (e e) (s s) (kp kp) (hash hash) (i i))
  (non-orig (invk hash) (privk e) (privk s))
  (uniq-orig nv)
  (traces
    ((recv (enc s r a ns (pubk v)))
      (recv (enc "cert" a i e (privk e)))
      (send (enc nv j v ns m (pubk s)))
      (recv
        (enc kp s jo m p
          (enc
            (enc "hash" (enc "hash" a v r nv j jo hash) m p hash)
            (invk i)) (pubk v)))))
  (label 61)
  (unrealized (0 1) (0 3))
  (comment "1 in cohort - 1 not yet seen"))
\end{verbatim}

$$\xymatrix{
\txt{\strut verifier}&\txt{\strut epca}&\txt{\strut server}&\txt{\strut attester}\\
&\bullet\ar[ldd]&\bullet\ar@{=>}[d]&\\
\bullet\ar@{=>}[d]&&\bullet\ar@{=>}[dd]\ar[ll]&\\
\bullet\ar@{=>}[d]&&&\\
\bullet\ar@{=>}[dd]\ar[rr]&&\bullet\ar@{=>}[d]&\\
&&\bullet\ar[r]&\bullet\ar@{=>}[d]\\
\bullet&&&\bullet\ar[lll]}
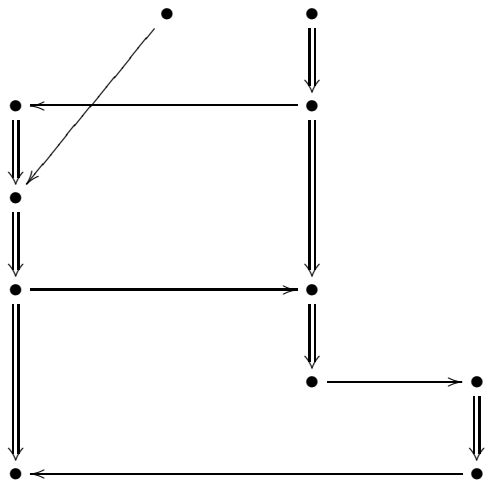
$$

\begin{verbatim}
(defskeleton caves
  (vars (r m p j jo r-0 text) (ns nv data) (e s a v a-0 s-0 name)
    (kp k kp-0 skey) (hash i akey))
  (defstrand verifier 4 (r r) (m m) (p p) (j j) (jo jo) (ns ns)
    (nv nv) (a a) (v v) (e e) (s s) (kp kp) (hash hash) (i i))
  (defstrand epca 1 (a a) (e e) (i i))
  (defstrand server 4 (r r-0) (m m) (j j) (ns ns) (nv nv) (a a-0)
    (v v) (s s) (k k))
  (defstrand attester 2 (r r) (m m) (p p) (j j) (jo jo) (nv nv)
    (a a) (v v) (s s-0) (kp kp-0) (hash hash) (i i))
  (precedes ((0 2) (2 2)) ((1 0) (0 1)) ((2 1) (0 0))
    ((2 3) (3 0)) ((3 1) (0 3)))
  (non-orig (invk hash) (invk i) (privk e) (privk s))
  (uniq-orig ns nv kp-0)
  (operation nonce-test (added-strand server 4) nv (3 0)
    (enc nv j v ns m (pubk s)))
  (traces
    ((recv (enc s r a ns (pubk v)))
      (recv (enc "cert" a i e (privk e)))
      (send (enc nv j v ns m (pubk s)))
      (recv
        (enc kp s jo m p
          (enc
            (enc "hash" (enc "hash" a v r nv j jo hash) m p hash)
            (invk i)) (pubk v))))
    ((send (enc "cert" a i e (privk e))))
    ((recv (enc r-0 a-0 k (pubk s)))
      (send (enc s r-0 a-0 ns (pubk v)))
      (recv (enc nv j v ns m (pubk s))) (send (enc nv j v m k)))
    ((recv (enc s-0 nv j v m r (ltk a a)))
      (send
        (enc kp-0 nv
          (enc kp-0 s-0 jo m p
            (enc
              (enc "hash" (enc "hash" a v r nv j jo hash) m p
                hash) (invk i)) (pubk v)) (ltk a a)))))
  (label 65)
  (parent 61)
  (unrealized)
  (shape)
  (annotations ((0 1) v (says e (id a i))) ((0 2) v (ask r a j m))
    ((0 3) v (says a (meas i nv j jo m p))) ((1 0) e (id a i))
    ((2 1) s (verifier v))
    ((3 1) a (and (verifier v) (meas i nv j jo m p))))
  (obligations
    ((0 1) v
      (implies (says e (id a i)) (says s (verifier v))
        (says e (id a i))))
    ((0 3) v
      (implies (ask r a j m) (says e (id a i))
        (says s (verifier v))
        (says a (and (verifier v) (meas i nv j jo m p)))
        (says a (meas i nv j jo m p))))))
\end{verbatim}

\subsection{Scenario}

$$\xymatrix{
\txt{\strut attester}\\
\bullet\ar@{=>}[d]\\
\bullet}$$

\begin{verbatim}
(defskeleton caves
  (vars (r m p j jo text) (nv data) (a v s name) (kp skey)
    (hash i akey))
  (defstrand attester 2 (r r) (m m) (p p) (j j) (jo jo) (nv nv)
    (a a) (v v) (s s) (kp kp) (hash hash) (i i))
  (non-orig (ltk a a) (invk hash) (privk v))
  (uniq-orig kp)
  (traces
    ((recv (enc s nv j v m r (ltk a a)))
      (send
        (enc kp nv
          (enc kp s jo m p
            (enc
              (enc "hash" (enc "hash" a v r nv j jo hash) m p
                hash) (invk i)) (pubk v)) (ltk a a)))))
  (label 66)
  (unrealized (0 0))
  (comment "1 in cohort - 1 not yet seen"))
\end{verbatim}

$$\xymatrix{
\txt{\strut attester}&\txt{\strut client}\\
&\bullet\ar@{=>}[d]\\
&\bullet\ar@{=>}[d]\\
\bullet\ar@{=>}[d]&\bullet\ar[l]\\
\bullet&}
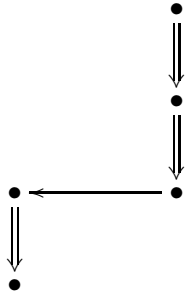
$$

\begin{verbatim}
(defskeleton caves
  (vars (r m p j jo text) (nv data) (a v s c name) (kp k skey)
    (hash i akey))
  (defstrand attester 2 (r r) (m m) (p p) (j j) (jo jo) (nv nv)
    (a a) (v v) (s s) (kp kp) (hash hash) (i i))
  (defstrand client 3 (r r) (m m) (j j) (nv nv) (c c) (a a) (v v)
    (s s) (k k))
  (precedes ((1 2) (0 0)))
  (non-orig (ltk a a) (invk hash) (privk v))
  (uniq-orig kp k)
  (operation encryption-test (added-strand client 3)
    (enc s nv j v m r (ltk a a)) (0 0))
  (traces
    ((recv (enc s nv j v m r (ltk a a)))
      (send
        (enc kp nv
          (enc kp s jo m p
            (enc
              (enc "hash" (enc "hash" a v r nv j jo hash) m p
                hash) (invk i)) (pubk v)) (ltk a a))))
    ((send (cat c (enc r a k (pubk s)))) (recv (enc nv j v m k))
      (send (enc s nv j v m r (ltk a a)))))
  (label 67)
  (parent 66)
  (unrealized)
  (shape)
  (annotations ((0 1) a (and (verifier v) (meas i nv j jo m p))))
  (obligations))
\end{verbatim}

\subsection{Scenario}

$$\xymatrix{
\txt{\strut attester}\\
\bullet\ar@{=>}[d]\\
\bullet}$$

\begin{verbatim}
(defskeleton caves
  (vars (r m p j jo text) (nv data) (a v s name) (kp skey)
    (hash i akey))
  (defstrand attester 2 (r r) (m m) (p p) (j j) (jo jo) (nv nv)
    (a a) (v v) (s s) (kp kp) (hash hash) (i i))
  (non-orig (invk hash) (privk v))
  (uniq-orig kp)
  (traces
    ((recv (enc s nv j v m r (ltk a a)))
      (send
        (enc kp nv
          (enc kp s jo m p
            (enc
              (enc "hash" (enc "hash" a v r nv j jo hash) m p
                hash) (invk i)) (pubk v)) (ltk a a)))))
  (label 68)
  (unrealized)
  (shape)
  (annotations ((0 1) a (and (verifier v) (meas i nv j jo m p))))
  (obligations))
\end{verbatim}

\subsection{Scenario}

$$\xymatrix{
\txt{\strut attester}&\txt{\strut }\\
\bullet\ar@{=>}[d]&\bullet\ar@{=>}[d]\\
\bullet&\bullet}
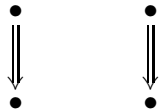
$$

\begin{verbatim}
(defskeleton caves
  (vars (jo r m p j text) (nv data) (a v s name) (kp skey)
    (hash i akey))
  (defstrand attester 2 (r r) (m m) (p p) (j j) (jo jo) (nv nv)
    (a a) (v v) (s s) (kp kp) (hash hash) (i i))
  (deflistener jo)
  (non-orig (invk hash) (privk v))
  (uniq-orig jo kp)
  (traces
    ((recv (enc s nv j v m r (ltk a a)))
      (send
        (enc kp nv
          (enc kp s jo m p
            (enc
              (enc "hash" (enc "hash" a v r nv j jo hash) m p
                hash) (invk i)) (pubk v)) (ltk a a))))
    ((recv jo) (send jo)))
  (label 69)
  (unrealized (1 0)))
\end{verbatim}

\subsection{Scenario}

$$\xymatrix{
\txt{\strut attester}&\txt{\strut }\\
\bullet\ar@{=>}[d]&\bullet\ar@{=>}[d]\\
\bullet&\bullet}
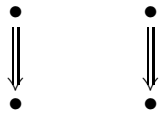
$$

\begin{verbatim}
(defskeleton caves
  (vars (p r m j jo text) (nv data) (a v s name) (kp skey)
    (hash i akey))
  (defstrand attester 2 (r r) (m m) (p p) (j j) (jo jo) (nv nv)
    (a a) (v v) (s s) (kp kp) (hash hash) (i i))
  (deflistener p)
  (non-orig (invk hash) (privk v))
  (uniq-orig p kp)
  (traces
    ((recv (enc s nv j v m r (ltk a a)))
      (send
        (enc kp nv
          (enc kp s jo m p
            (enc
              (enc "hash" (enc "hash" a v r nv j jo hash) m p
                hash) (invk i)) (pubk v)) (ltk a a))))
    ((recv p) (send p)))
  (label 71)
  (unrealized (1 0)))
\end{verbatim}

\subsection{Scenario}

$$\xymatrix{
\txt{\strut server}\\
\bullet\ar@{=>}[d]\\
\bullet\ar@{=>}[d]\\
\bullet\ar@{=>}[d]\\
\bullet\ar@{=>}[d]\\
\bullet\ar@{=>}[d]\\
\bullet\ar@{=>}[d]\\
\bullet\ar@{=>}[d]\\
\bullet}$$

\begin{verbatim}
(defskeleton caves
  (vars (b mesg) (r m j d text) (ns nv data) (v s a name)
    (k kp skey))
  (defstrand server 8 (b b) (r r) (m m) (j j) (d d) (ns ns)
    (nv nv) (a a) (v v) (s s) (k k) (kp kp))
  (non-orig (privk v) (privk s))
  (uniq-orig ns)
  (traces
    ((recv (enc r a k (pubk s))) (send (enc s r a ns (pubk v)))
      (recv (enc nv j v ns m (pubk s))) (send (enc nv j v m k))
      (recv b) (send b) (recv (enc "valid" kp ns (pubk s)))
      (send (enc "data" d kp))))
  (label 73)
  (unrealized (0 2) (0 6))
  (comment "1 in cohort - 1 not yet seen"))
\end{verbatim}

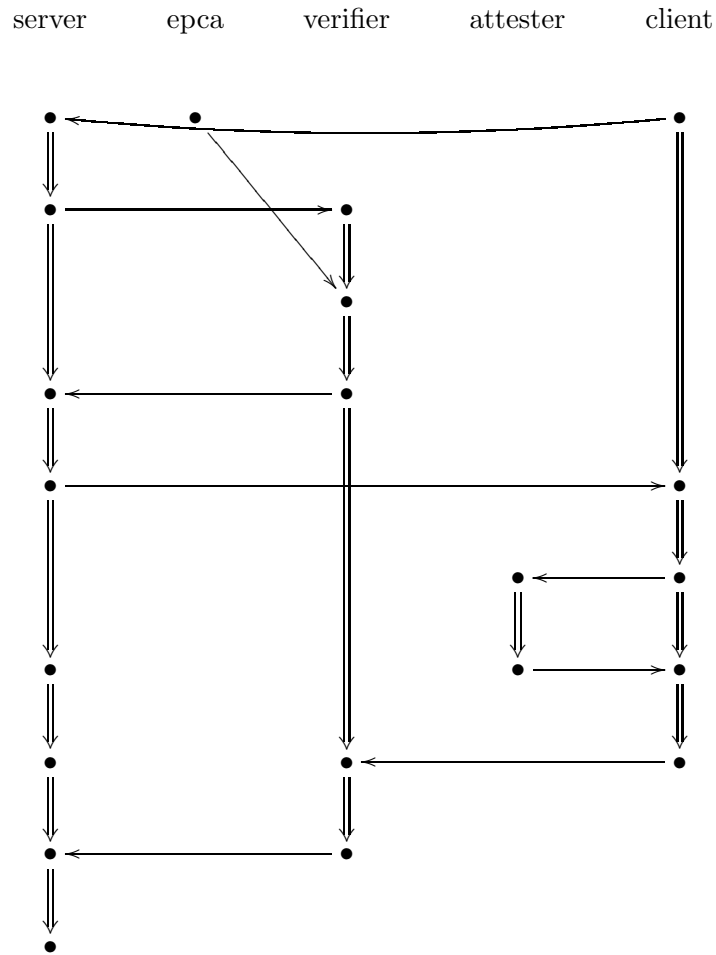
$$\xymatrix{
\txt{\strut server}&\txt{\strut epca}&\txt{\strut verifier}&\txt{\strut attester}&\txt{\strut client}\\
\bullet\ar@{=>}[d]&\bullet\ar[rdd]&&&\bullet\ar@{=>}[dddd]\ar@/^/[llll]\\
\bullet\ar@{=>}[dd]\ar[rr]&&\bullet\ar@{=>}[d]&&\\
&&\bullet\ar@{=>}[d]&&\\
\bullet\ar@{=>}[d]&&\bullet\ar@{=>}[dddd]\ar[ll]&&\\
\bullet\ar@{=>}[dd]\ar[rrrr]&&&&\bullet\ar@{=>}[d]\\
&&&\bullet\ar@{=>}[d]&\bullet\ar@{=>}[d]\ar[l]\\
\bullet\ar@{=>}[d]&&&\bullet\ar[r]&\bullet\ar@{=>}[d]\\
\bullet\ar@{=>}[d]&&\bullet\ar@{=>}[d]&&\bullet\ar[ll]\\
\bullet\ar@{=>}[d]&&\bullet\ar[ll]&&\\
\bullet&&&&}$$

\begin{verbatim}
(defskeleton caves
  (vars (b mesg) (d m j p jo r text) (ns nv data) (v a e c s name)
    (k kp skey) (hash i akey))
  (defstrand server 8 (b b) (r r) (m m) (j j) (d d) (ns ns)
    (nv nv) (a a) (v v) (s s) (k k) (kp kp))
  (defstrand epca 1 (a a) (e e) (i i))
  (defstrand verifier 5 (r r) (m m) (p p) (j j) (jo jo) (ns ns)
    (nv nv) (a a) (v v) (e e) (s s) (kp kp) (hash hash) (i i))
  (defstrand attester 2 (r r) (m m) (p p) (j j) (jo jo) (nv nv)
    (a a) (v v) (s s) (kp kp) (hash hash) (i i))
  (defstrand client 5
    (b
      (enc kp s jo m p
        (enc (enc "hash" (enc "hash" a v r nv j jo hash) m p hash)
          (invk i)) (pubk v))) (r r) (m m) (j j) (nv nv) (c c)
    (a a) (v v) (s s) (k k) (kp kp))
  (precedes ((0 1) (2 0)) ((0 3) (4 1)) ((1 0) (2 1))
    ((2 2) (0 2)) ((2 4) (0 6)) ((3 1) (4 3)) ((4 0) (0 0))
    ((4 2) (3 0)) ((4 4) (2 3)))
  (non-orig (ltk a a) (invk hash) (invk i) (privk v) (privk e)
    (privk s))
  (uniq-orig ns nv k kp)
  (operation nonce-test (contracted (kp-0 kp)) ns (0 6)
    (enc "valid" kp ns (pubk s)) (enc nv j v ns m (pubk s))
    (enc s r a ns (pubk v)))
  (traces
    ((recv (enc r a k (pubk s))) (send (enc s r a ns (pubk v)))
      (recv (enc nv j v ns m (pubk s))) (send (enc nv j v m k))
      (recv b) (send b) (recv (enc "valid" kp ns (pubk s)))
      (send (enc "data" d kp)))
    ((send (enc "cert" a i e (privk e))))
    ((recv (enc s r a ns (pubk v)))
      (recv (enc "cert" a i e (privk e)))
      (send (enc nv j v ns m (pubk s)))
      (recv
        (enc kp s jo m p
          (enc
            (enc "hash" (enc "hash" a v r nv j jo hash) m p hash)
            (invk i)) (pubk v)))
      (send (enc "valid" kp ns (pubk s))))
    ((recv (enc s nv j v m r (ltk a a)))
      (send
        (enc kp nv
          (enc kp s jo m p
            (enc
              (enc "hash" (enc "hash" a v r nv j jo hash) m p
                hash) (invk i)) (pubk v)) (ltk a a))))
    ((send (cat c (enc r a k (pubk s)))) (recv (enc nv j v m k))
      (send (enc s nv j v m r (ltk a a)))
      (recv
        (enc kp nv
          (enc kp s jo m p
            (enc
              (enc "hash" (enc "hash" a v r nv j jo hash) m p
                hash) (invk i)) (pubk v)) (ltk a a)))
      (send
        (enc kp s jo m p
          (enc
            (enc "hash" (enc "hash" a v r nv j jo hash) m p hash)
            (invk i)) (pubk v)))))
  (label 96)
  (parent 73)
  (unrealized)
  (shape)
  (annotations ((0 1) s (verifier v))
    ((0 6) s (says v (approved r a nv)))
    ((0 7) s (and (approved r a nv) (resource r d)))
    ((1 0) e (id a i)) ((2 1) v (says e (id a i)))
    ((2 2) v (ask r a j m))
    ((2 3) v (says a (meas i nv j jo m p)))
    ((2 4) v (approved r a nv))
    ((3 1) a (and (verifier v) (meas i nv j jo m p))))
  (obligations
    ((0 6) s
      (implies (verifier v) (says e (id a i))
        (says v (ask r a j m)) (says v (approved r a nv))
        (says a (and (verifier v) (meas i nv j jo m p)))
        (says v (approved r a nv))))
    ((2 1) v
      (implies (says s (verifier v)) (says e (id a i))
        (says e (id a i))))
    ((2 3) v
      (implies (says s (verifier v)) (says e (id a i))
        (ask r a j m)
        (says a (and (verifier v) (meas i nv j jo m p)))
        (says a (meas i nv j jo m p))))))
\end{verbatim}

\subsection{Scenario}

$$\xymatrix{
\txt{\strut server}&\txt{\strut }\\
\bullet\ar@{=>}[d]&\bullet\ar@{=>}[d]\\
\bullet\ar@{=>}[d]&\bullet\\
\bullet\ar@{=>}[d]&\\
\bullet\ar@{=>}[d]&\\
\bullet\ar@{=>}[d]&\\
\bullet\ar@{=>}[d]&\\
\bullet\ar@{=>}[d]&\\
\bullet&}
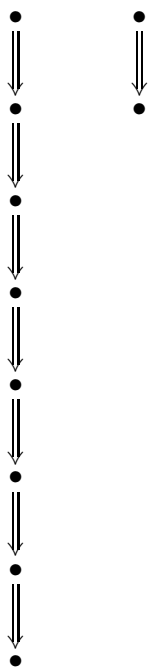
$$

\begin{verbatim}
(defskeleton caves
  (vars (b mesg) (d r m j text) (ns nv data) (v s a name)
    (k kp skey))
  (defstrand server 8 (b b) (r r) (m m) (j j) (d d) (ns ns)
    (nv nv) (a a) (v v) (s s) (k k) (kp kp))
  (deflistener d)
  (non-orig (privk v) (privk s))
  (uniq-orig d ns)
  (traces
    ((recv (enc r a k (pubk s))) (send (enc s r a ns (pubk v)))
      (recv (enc nv j v ns m (pubk s))) (send (enc nv j v m k))
      (recv b) (send b) (recv (enc "valid" kp ns (pubk s)))
      (send (enc "data" d kp))) ((recv d) (send d)))
  (label 107)
  (unrealized (0 2) (0 6) (1 0)))
\end{verbatim}

\subsection{Scenario}

$$\xymatrix{
\txt{\strut client}\\
\bullet\ar@{=>}[d]\\
\bullet\ar@{=>}[d]\\
\bullet\ar@{=>}[d]\\
\bullet\ar@{=>}[d]\\
\bullet\ar@{=>}[d]\\
\bullet}$$

\begin{verbatim}
(defskeleton caves
  (vars (b mesg) (r m j d text) (nv data) (a v s c name)
    (k kp skey))
  (defstrand client 6 (b b) (r r) (m m) (j j) (d d) (nv nv) (c c)
    (a a) (v v) (s s) (k k) (kp kp))
  (non-orig (ltk a a) (privk v) (privk s))
  (uniq-orig k)
  (traces
    ((send (cat c (enc r a k (pubk s)))) (recv (enc nv j v m k))
      (send (enc s nv j v m r (ltk a a)))
      (recv (enc kp nv b (ltk a a))) (send b)
      (recv (enc "data" d kp))))
  (label 157)
  (unrealized (0 1) (0 3))
  (comment "2 in cohort - 2 not yet seen"))
\end{verbatim}

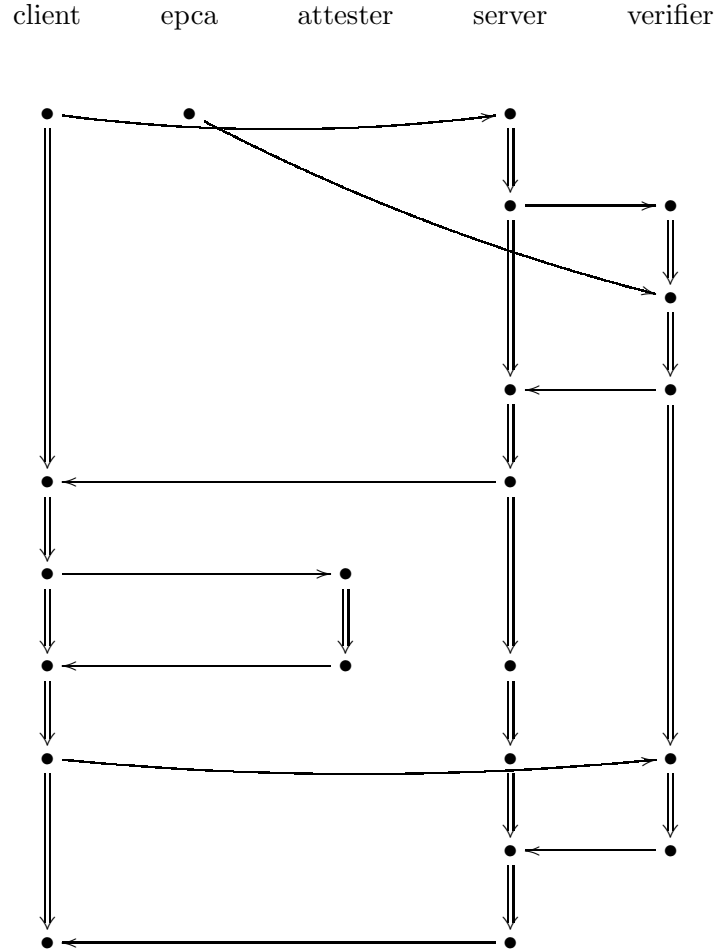
$$\xymatrix{
\txt{\strut client}&\txt{\strut epca}&\txt{\strut attester}&\txt{\strut server}&\txt{\strut verifier}\\
\bullet\ar@{=>}[dddd]\ar@/_/[rrr]&\bullet\ar@/_/[rrrdd]&&\bullet\ar@{=>}[d]&\\
&&&\bullet\ar@{=>}[dd]\ar[r]&\bullet\ar@{=>}[d]\\
&&&&\bullet\ar@{=>}[d]\\
&&&\bullet\ar@{=>}[d]&\bullet\ar@{=>}[dddd]\ar[l]\\
\bullet\ar@{=>}[d]&&&\bullet\ar@{=>}[dd]\ar[lll]&\\
\bullet\ar@{=>}[d]\ar[rr]&&\bullet\ar@{=>}[d]&&\\
\bullet\ar@{=>}[d]&&\bullet\ar[ll]&\bullet\ar@{=>}[d]&\\
\bullet\ar@{=>}[dd]\ar@/_/[rrrr]&&&\bullet\ar@{=>}[d]&\bullet\ar@{=>}[d]\\
&&&\bullet\ar@{=>}[d]&\bullet\ar[l]\\
\bullet&&&\bullet\ar[lll]&}$$

\begin{verbatim}
(defskeleton caves
  (vars (b mesg) (d p jo r m j text) (nv ns data) (s c a v e name)
    (kp k skey) (hash i akey))
  (defstrand client 6
    (b
      (enc kp s jo m p
        (enc (enc "hash" (enc "hash" a v r nv j jo hash) m p hash)
          (invk i)) (pubk v))) (r r) (m m) (j j) (d d) (nv nv)
    (c c) (a a) (v v) (s s) (k k) (kp kp))
  (defstrand epca 1 (a a) (e e) (i i))
  (defstrand attester 2 (r r) (m m) (p p) (j j) (jo jo) (nv nv)
    (a a) (v v) (s s) (kp kp) (hash hash) (i i))
  (defstrand server 8 (b b) (r r) (m m) (j j) (d d) (ns ns)
    (nv nv) (a a) (v v) (s s) (k k) (kp kp))
  (defstrand verifier 5 (r r) (m m) (p p) (j j) (jo jo) (ns ns)
    (nv nv) (a a) (v v) (e e) (s s) (kp kp) (hash hash) (i i))
  (precedes ((0 0) (3 0)) ((0 2) (2 0)) ((0 4) (4 3))
    ((1 0) (4 1)) ((2 1) (0 3)) ((3 1) (4 0)) ((3 3) (0 1))
    ((3 7) (0 5)) ((4 2) (3 2)) ((4 4) (3 6)))
  (non-orig (ltk a a) (invk hash) (invk i) (privk s) (privk v)
    (privk e))
  (uniq-orig nv ns kp k)
  (operation nonce-test
    (contracted (a-0 a) (v-0 v) (k-0 k) (r-0 r) (m-0 m) (j-0 j)
      (nv-0 nv) (s-0 s) (ns-0 ns)) kp (4 6)
    (enc "valid" kp ns (pubk s))
    (enc kp nv
      (enc kp s jo m p
        (enc (enc "hash" (enc "hash" a v r nv j jo hash) m p hash)
          (invk i)) (pubk v)) (ltk a a))
    (enc kp s jo m p
      (enc (enc "hash" (enc "hash" a v r nv j jo hash) m p hash)
        (invk i)) (pubk v)))
  (traces
    ((send (cat c (enc r a k (pubk s)))) (recv (enc nv j v m k))
      (send (enc s nv j v m r (ltk a a)))
      (recv
        (enc kp nv
          (enc kp s jo m p
            (enc
              (enc "hash" (enc "hash" a v r nv j jo hash) m p
                hash) (invk i)) (pubk v)) (ltk a a)))
      (send
        (enc kp s jo m p
          (enc
            (enc "hash" (enc "hash" a v r nv j jo hash) m p hash)
            (invk i)) (pubk v))) (recv (enc "data" d kp)))
    ((send (enc "cert" a i e (privk e))))
    ((recv (enc s nv j v m r (ltk a a)))
      (send
        (enc kp nv
          (enc kp s jo m p
            (enc
              (enc "hash" (enc "hash" a v r nv j jo hash) m p
                hash) (invk i)) (pubk v)) (ltk a a))))
    ((recv (enc r a k (pubk s))) (send (enc s r a ns (pubk v)))
      (recv (enc nv j v ns m (pubk s))) (send (enc nv j v m k))
      (recv b) (send b) (recv (enc "valid" kp ns (pubk s)))
      (send (enc "data" d kp)))
    ((recv (enc s r a ns (pubk v)))
      (recv (enc "cert" a i e (privk e)))
      (send (enc nv j v ns m (pubk s)))
      (recv
        (enc kp s jo m p
          (enc
            (enc "hash" (enc "hash" a v r nv j jo hash) m p hash)
            (invk i)) (pubk v)))
      (send (enc "valid" kp ns (pubk s)))))
  (label 210)
  (parent 157)
  (unrealized)
  (shape)
  (annotations ((0 5) c (says s (resource r d)))
    ((1 0) e (id a i))
    ((2 1) a (and (verifier v) (meas i nv j jo m p)))
    ((3 1) s (verifier v)) ((3 6) s (says v (approved r a nv)))
    ((3 7) s (and (approved r a nv) (resource r d)))
    ((4 1) v (says e (id a i))) ((4 2) v (ask r a j m))
    ((4 3) v (says a (meas i nv j jo m p)))
    ((4 4) v (approved r a nv)))
  (obligations
    ((0 5) c
      (implies (says e (id a i))
        (says a (and (verifier v) (meas i nv j jo m p)))
        (says s (verifier v))
        (says s (and (approved r a nv) (resource r d)))
        (says v (ask r a j m)) (says v (approved r a nv))
        (says s (resource r d))))
    ((3 6) s
      (implies (says e (id a i))
        (says a (and (verifier v) (meas i nv j jo m p)))
        (verifier v) (says v (ask r a j m))
        (says v (approved r a nv)) (says v (approved r a nv))))
    ((4 1) v
      (implies (says e (id a i)) (says s (verifier v))
        (says e (id a i))))
    ((4 3) v
      (implies (says e (id a i))
        (says a (and (verifier v) (meas i nv j jo m p)))
        (says s (verifier v)) (ask r a j m)
        (says a (meas i nv j jo m p))))))
\end{verbatim}
\endgroup
\fi

\end{document}